\documentclass[%
 reprint,
 amsmath,amssymb,
 aps,
]{revtex4-1}

\usepackage{graphicx}% Include figure files
\usepackage{dcolumn}% Align table columns on decimal point
\usepackage{bm}% bold math

\usepackage[linktocpage=true,
  colorlinks=true, 
  pdfborder={0 0 0},
  linkcolor=blue,
  citecolor=red,
  filecolor=yellow,
  urlcolor=blue,
  bookmarks,
  pdfauthor={},
]{hyperref}

\begin{document}

\preprint{APS/123-QED}

\title[GF_confined_two_electrons]{
One-particle Green's function of interacting two electrons using \\
analytic solutions for a three-body problem: \\
comparison with exact Kohn--Sham system
}

\author{Taichi Kosugi}
\affiliation{
Department of Physics,
University of Tokyo,
Tokyo 113-0033,
Japan 
}

\author{Yu-ichiro Matsushita}
\affiliation{
Laboratory for Materials and Structures,
Institute of Innovative Research,
Tokyo Institute of Technology,
Yokohama 226-8503,
Japan
}

\date{\today}

\begin{abstract}
For a three-electron system with finite-strength interactions confined to a one-dimensional harmonic trap,
we solve the Schr\"odinger equation analytically to obtain the exact solutions,
from which we construct explicitly the simultaneous eigenstates of
the energy and total spin for the first time.
The solutions for the three-electron system allow us to derive analytic expressions for the exact one-particle Green's function (GF) for the corresponding two-electron system.
We calculate the GF in frequency domain to examine systematically its behavior depending on the electronic interactions. 
We also compare the pole structure of non-interacting GF using the exact Kohn--Sham (KS) potential with that of the exact GF
to find that the discrepancy of the energy gap between the KS system and the original system is larger for a stronger interaction.
We perform numerical examination on the behavior of GFs in real space to demonstrate that the exact and KS GFs can have shapes quite different from each other.
Our simple model will help to understand generic characteristics of interacting GFs.
\end{abstract}

\maketitle

\section{introduction}

An interacting two-electron system confined to a three-dimensional harmonic trap, called a harmonium\cite{bib:2817,bib:906,bib:3926,bib:4475},
has been studied intensively since the Schr\"odinger equation for this real-space model is analytically solvable\cite{bib:906} for a particular set of the confinement strength.
This model is interesting not only from a theoretical viewpoint,
but also from a practical viewpoint because of the recent progress of fabrication techniques to realize artificial many-electron systems such as quantum dots and confined ultracold atoms.\cite{bib:4492}
Even such simple systems are known to provide insights into electronic correlation effects that cannot be captured in single-particle picture.

For quantitative description of realistic systems, on the other hand,
electronic-structure calculations via solving the Kohn--Sham (KS) equations
based on the density functional theory (DFT)\cite{bib:76,bib:77} have been successful for a large part of target materials.
It is known, however, that DFT calculations often fail to explain the properties of strongly correlated systems even qualitatively.
To remedy this drawback, various approaches have been proposed.
There exist such approaches based on the Green's function (GF) theory, including $GW$ approaches.\cite{bib:GW1,bib:GW2,bib:GW3}
The GF-based prescriptions often use the non-interacting electronic states obtained in DFT calculations as reference states for the construction of interacting GFs.
Since the foundation of DFT ensures the correctness only of the calculated electron density and total energy of an interacting system
even if we knew the exact exchange correlation functional,
it is useful to have a simple many-electron model which enables one to calculate the exact GF and compare it with the one constructed via the DFT calculation.

Photoelectron spectroscopy has become an active research field today.
Measurements of photoelectron spectra performed in various ways such as angle-resolved photoemission spectroscopy (ARPES) for clarifying the properties of materials.
The measured spectra are often explained under a certain assumption via the one-particle GF of a many-electron system.\cite{bib:4070,bib:4165,bib:pw_unfolding}
The clear understanding of the characteristics of GFs is thus important not only for theoretical studies,
but also for practical studies in material science.
The calculation of GFs in the context of explicitly correlated electronic-structure calculations for realistic systems has been drawing attention recently.\cite{bib:4473,bib:4483,bib:GF_CCSD_and_FCI_for_light_atoms}

Given the progresses on the theoretical and experimental studies outlined above,
one notices that it is worth finding a new exactly solvable many-electron model in real space
and proposing a minimal model for an exact interacting one-particle GF. 
In the present study,
we obtain the exact solutions of the Schr\"odinger equation for a three-electron system with finite-strength interactions confined to a one-dimensional trap,
from which we construct explicitly the simultaneous eigenstates of
the energy and total spin for the first time, to our best knowledge.
In addition, we provide the exact one-particle GF of the corresponding two-electron system and compare it with that in the KS system.
Analytic solutions for an arbitrary number of confined spin-$1/2$ fermions with delta-type interactions in the strong-interaction limit have been obtained by employing a group-theoretic approach.\cite{bib:4493}
Numerical solutions for such fermions with particle numbers more than two have also been reported for finite-strength interactions.\cite{bib:4494}
There exist various theoretical studies on confined interacting particles in the context of one-dimensional Fermi gases.\cite{bib:4492}

This paper is organized as follows.
In Section \ref{section:exact_solutions},
we provide the analytic solutions of the Schr\"odinger equation for the interacting three-electron system,
from which we construct the eigenstates of the energy and total spin.
In Section \ref{section:GF},
we derive the exact expressions for the one-particle GF of the two-electron system and perform numerical calculations to examine the behavior of the GFs.
In Section \ref{section:conclusions},
the conclusions are provided.

\section{exact solutions of Schr\"odinger equations}
\label{section:exact_solutions}

\subsection{One-electron system}

The Hamiltonian for an electron confined to a one-dimensional harmonic trap 
$V_0 (x_1) = m_e \omega_0^2 x_1^2/2$
with its strength $\omega_0$ is
$
H^{\mathrm{HO}} (x_1; m_e, \omega_0)
=
-1/(2 m_e)
\partial^2/\partial x_1^2
+
V_0 (x_1)
.
$
$m_e$ is the electron mass.
The orthonormalized energy eigenfunction is given by
$
\psi_n (x_1)
=
\psi^{\mathrm{HO}}_{n} (\overline{x}_1; \ell)
\equiv
1/ [ \ell^{1/2} (2 \pi)^{1/4} \sqrt{2^n n !} ]
\exp (-\overline{x}_1^2/4)
H_n ( \overline{x}_1/\sqrt{2})
,
$
where
$\ell \equiv 1/\sqrt{2 m_e \omega_0}$
is the typical length scale and
$\overline{x}_1 \equiv x_1/\ell$ is the dimensionless coordinate.
$H_n (x_1)$ is the Hermite polynomial.
The energy eigenvalue for the quantum number $n$ is
$E^{(1)}_n = \omega_0 ( n + 1/2 )$,
as found in textbooks of quantum mechanics.
The corresponding wave function (WF) for an electron is given by
$
\Psi_n^{S_z} (x_1, s_1)
=
\psi^{\mathrm{HO}}_{n} (x_1/\ell; \ell)
\chi^{S = 1/2, S_z} (s_1)
.
$
$\chi^{S = 1/2, S_z}$ is the spin WF with
$\chi^{1/2, 1/2} (s_1) = \alpha (s_1)$
and
$\chi^{1/2, -1/2} (s_1) = \beta (s_1)$.

\subsection{Two-electron system}

The Hamiltonian for the confined interacting two electrons considered in the present work is
\begin{gather}
	H^{(2)}
    =
		H^{\mathrm{HO}} (x_1; m_e, \omega_0)
        +
		H^{\mathrm{HO}} (x_2; m_e, \omega_0)
	\nonumber \\
    	+ v (x_1 - x_2)
	,
	\label{def_H2}        
\end{gather}
where
\begin{gather}
	v (x)
    =
    	-
        \frac{m_e \omega_0^2 \Lambda}{2}
        x^2
\end{gather}
is the repulsive interaction between the electrons.
The dimensionless parameter $0 \leqq \Lambda \leqq 1/3$ measures the strength of interaction.
This model has been studied intensively.\cite{bib:3964,bib:2826,bib:2825,bib:4276}
The variable transformation for the center-of-mass coordinate
$x_+ \equiv (x_1 + x_2)/2$
and the scaled relative coordinate
$x_- \equiv (x_1 - x_2)/\sqrt{2}$
decouples the interacting Hamiltonian into two Hamiltonians for independent harmonic oscillators as
$
H^{(2)}
=
H^{\mathrm{HO}} (x_+; M_2, \omega_0)
+
H^{\mathrm{HO}} (x_-; m_e, \omega_{\mathrm{r}})
,
$
where
$M_2 \equiv 2 m_e$ and
$
\omega_{\mathrm{r}}
\equiv
\sqrt{1 - 2 \Lambda } \omega_0
\equiv
\lambda_2 \omega_0
$.
The solution for the center-of-mass motion is
$
\psi_{\mathrm{c} n_\mathrm{c}} (x_+)
=
\psi^{\mathrm{HO}}_{n_{\mathrm{c}}}
(\sqrt{2} x_+/\ell; \ell/\sqrt{2})
$
and that for the relative motion is
$
\psi_{\mathrm{r} n_\mathrm{r}} (x_-)
=
\psi^{\mathrm{HO}}_{n_{\mathrm{r}}}
(\sqrt{\lambda_2} x_-/\ell; \ell/\sqrt{\lambda_2})
$.
With the quantum numbers $n_{\mathrm{c}}$ and $n_{\mathrm{r}}$,
the energy eigenvalue for the whole system is given simply by the sum of two energy eigenvalues for the two oscillators:
\begin{gather}
	E^{(2)}_{n_\mathrm{c} n_\mathrm{r}} 
	=
		\omega_0  \left( n_\mathrm{c} + \frac{1}{2} \right)
		+
        \lambda_2
		\omega_0 \left( n_\mathrm{r} + \frac{1}{2} \right)
	,
	\label{energy_eval_2}
\end{gather}
whose corresponding normalized spatial WF is
\begin{gather}
   	\Psi_{n_{\mathrm{c}} n_{\mathrm{r}} } (x_1, x_2)
	=
		2^{1/4}
		\psi_{\mathrm{c} n_{\mathrm{c}}} (x_+) 
		\psi_{\mathrm{r} n_{\mathrm{r}}} (x_-) 
		.
	\label{harm_delta_two_el_wave_func}
\end{gather}
The Fermi statistics forces the two-electron WFs for the energy eigenstates to be in the following two forms:
\begin{gather}
	\Psi_{n_\mathrm{c} n_\mathrm{r}}^{S_z = 0}
    (x_1, s_1, x_2, s_2)
	=
		\Psi_{n_{\mathrm{c}} n_{\mathrm{r}} } (x_1, x_2)
		\chi^{0, 0} (s_1, s_2) 
	\label{WF_2e_even_nr}
\end{gather}
with an even $n_{\mathrm{r}}$ for a spin-singlet state and
\begin{gather}
	\Psi_{n_\mathrm{c} n_\mathrm{r}}^{S_z}
    (x_1, s_1, x_2, s_2)
	=
		\Psi_{n_{\mathrm{c}} n_{\mathrm{r}} } (x_1, x_2)
		\chi^{1, S_z} (s_1, s_2) 
	\label{WF_2e_odd_nr}
\end{gather}
with an odd $n_{\mathrm{r}}$ for spin-triplet states.
$\chi^{S, S_z} (s_1, s_2)$ is the spin WF for two electrons having the total spin $S$ and its $z$ component $S_z$.
The ground state is non-degenerate and its WF is given by
\begin{gather}
	\Psi_{\mathrm{gs}}^{(2)}
   	(x_1, s_1, x_2, s_2)
	=
		\Psi_{0 0}^{S = 0, S_z = 0}
    	(x_1, s_1, x_2, s_2)
	\label{WF2e_gs}
\end{gather}
with the energy eigenvalue
$
E_{\mathrm{gs}}^{(2)}
=
E_{0 0}^{(2)}
.
$

\subsection{Three-electron system}

\subsubsection{Variable transformations}

Calogero\cite{bib:3935} provided the exact analytic solutions for an interacting spinless three-particle system confined to a one-dimensional trap.
In the present study, we adopt his approach to obtain the solutions for a three-electron system necessary for deriving the expressions for the GF of the two-electron system.

The Hamiltonian for the interacting three-electron system we have to consider is
\begin{gather}
	H^{(3)}
    =
		H^{\mathrm{HO}} (x_1; m_e, \omega_0)
        +
		H^{\mathrm{HO}} (x_2; m_e, \omega_0)
	\nonumber \\
        +
		H^{\mathrm{HO}} (x_3; m_e, \omega_0)
	\nonumber \\
    	+
        v (x_1 - x_2)
        +
        v (x_2 - x_3)
        +
        v (x_3 - x_1)   	
	.
	\label{def_H3}        
\end{gather}
The energy eigenstates of this system are necessary for the calculation of GF for the two-electron system described above.
The interactions considered by Calogero\cite{bib:3935} was inverse-quadratic interactions in addition to those present in eq. (\ref{def_H3}).

It is appropriate to perform the two successive variable transformations.
The first one is $(x_1, x_2, x_3) \to (X, x, y)$:
\begin{gather}
	X 
    \equiv
    	\frac{1}{3}
        (x_1 + x_2 + x_3)
	\label{X_in_x1_x2_x3}
	\\
    x 
    \equiv
    	\frac{1}{\sqrt{2}}
        (x_1 - x_2)
	\label{x_in_x1_x2}
	\\
    y 
    \equiv
    	\frac{1}{\sqrt{6}}
        (x_1 + x_2 - 2 x_3)
	,
	\label{y_in_x1_x2_x3}
\end{gather}
which are called the Jacobi coordinates.
The second one is $(X, x, y) \to (X, r, \phi)$:
\begin{gather}
	r 
    \equiv
    	\sqrt{x^2 + y^2}
	\label{def_r_in_x_y}
	\\
	\phi
    \equiv
    	\begin{cases}
    		\arctan \frac{x}{y} & (x \geqq 0) \\
    		\arctan \frac{x}{y} + \pi & (x < 0) \\
    	\end{cases}
	,
	\label{def_phi_in_x_y}
\end{gather}
where we define the range of $\arctan$ as $0 \leqq \arctan < \pi$.
$\phi$ is the angle between the $y$ axis and the line connecting the origin and $(x, y)$,
so that $x = r \sin \phi$ and $y = r \cos \phi$.
By using the transformed variables,
the Hamiltonian in eq. (\ref{def_H3}) are rewritten as
$H^{(3)} = H^{\mathrm{HO}} (X; M_3, \omega_0) + H_{\mathrm{rad}} (r, \phi)$,
where
\begin{gather}
    H_{\mathrm{rad}} (r, \phi)
	\equiv
    	-\frac{1}{2 m_e}
        \left[
			\frac{\partial^2}{\partial r^2}
   			+
        	\frac{1}{r}
			\frac{\partial}{\partial r}
   			-
        	\frac{L (\phi)}{r^2}
        \right]
        +
        \frac{m_e \omega_0^2 \lambda_3^2 }{2}
        r^2
	,
	\label{H3_for_r_phi}        
\end{gather}
$L (\phi) \equiv -\partial^2/ \partial \phi^2,
M_3 \equiv 3 m_e
$
, and
$\lambda_3 \equiv \sqrt{1 - 3 \Lambda}$.
For $\Lambda > 1/3$, such strong repulsive interactions do not allow the three electrons to form a bound state.

$\phi (x_1, x_2, x_3) \equiv \phi (123)$ is anti-symmetric under the exchange of $x_1$ and $x_2$
[see eqs. (\ref{x_in_x1_x2})-(\ref{def_phi_in_x_y})].
In addition,
as proved in Appendix \ref{appendix:proof_of_phi123},
$\phi (123)$ and those with permuted $x_1, x_2$, and $x_3$ are related via the following relations:
\begin{gather}
	\phi (231)
    =
    	\phi (123)
        +
        \frac{2 \pi}{3}
	, \,
	\phi (312)
    =
    	\phi (123)
        +
        \frac{4 \pi}{3}
	.
    \label{phi123_phi231_phi312_simple}
\end{gather}

\subsubsection{Spatial wave functions}

The Hamiltonian in eq. (\ref{H3_for_r_phi}) suggests the solution for the spatial WF of the form
$
\Psi^\pm (x_1, x_2, x_3)
=
\psi^{\mathrm{HO}}_k (\sqrt{3} X/\ell; \ell/\sqrt{3})
R (r)
\Phi_m^\pm (\phi)
$
, where
$\Phi_m^+ (\phi) = \cos (m \phi)/\sqrt{\pi}$ for $m = 0, 1, \dots$
and
$\Phi_m^- (\phi) = \sin (m \phi)/\sqrt{\pi}$ for $m = 1, 2, \dots$
.
Substitution of this WF into the time-independent Schr\"odinger equation
$H^{(3)} \Psi^\pm = E^{(3)} \Psi^\pm$
with the Hamiltonian in eq. (\ref{H3_for_r_phi}) leads to the following eigenvalue equation that has to be satisfied by $R (r)$:
\begin{gather}
	\left[
    	-\frac{1}{2 m_e}
        \left(
			\frac{d^2}{d r^2}
   			+
        	\frac{1}{r}
			\frac{d}{d r}
   			-
        	\frac{m^2}{r^2}
        \right)
        +
        \frac{ m_e \omega_0^2 \lambda_3^2}{2}
        r^2
	\right]
    R (r)
    =
    	E_{\mathrm{r}}
        R (r)
	\label{eigen_eq_R_r}
\end{gather}
$E_{\mathrm{r}} \equiv E^{(3)} - E_{\mathrm{c} k}$
and
$E_{\mathrm{c} k} \equiv \omega_0 (k + 1/2)$
is the contribution from the center-of-mass motion to the total energy.
We assume the decaying solution of the form
$R (r) = \overline{r}^m \exp (-\overline{r}^2/2) u (r)$
using the dimensionless coordinate
$\overline{r} \equiv \sqrt{\lambda_3/2} (r/\ell)$
and substitute it into eq. (\ref{eigen_eq_R_r})
to obtain the following equation:
\begin{gather}
	2
    \rho
    \frac{d^2 u}{d \rho^2}
	+
    2 (m + 1 - \rho)
    \frac{du}{d \rho}
    +
    \left( \frac{E_{\mathrm{r}}}{\lambda_3 \omega_0} - m - 1 \right)
    u
    =
    	0
	,
	\label{eigen_eq_u}
\end{gather}
where $\rho \equiv \overline{r}^2$.
In order for $u$ to be bounded when expressed in a series of $\rho$,
the series has to be truncated at a some order $n$.
This condition leads to the eigenvalue as
$E_{\mathrm{r} n m} = \lambda_3 \omega_0 (2 n + m + 1 )$.
%The equation in eq. (\ref{eigen_eq_u}) for the eigenfunction $u_{n m}$ belonging to $E_{\mathrm{r} n m}$ %thus reads
%\begin{gather}
%    \rho
%    \frac{d^2 u_{n m}}{d \rho^2}
%	+
%    (m + 1 - \rho)
%    \frac{d u_{n m}}{d \rho}
%    +
%    n
%    u_{n m}
%    =
%    	0
%	.        
%\end{gather}
%The solution of this equation is the associated Laguerre polynomial $L_n^m (\rho)$,
The solution of eq. (\ref{eigen_eq_u}) for the eigenfunction $u_{n m}$ belonging to
$E_{\mathrm{r} n m}$ is the associated Laguerre polynomial $L_n^m (\rho)$,
with which the solution of the radial equation in eq. (\ref{eigen_eq_R_r}) is then 
\begin{gather}
	R_{n m} (r)
    =
        \sqrt{ \frac{2 \sqrt{3} n !}{(n + m) !} }
		\frac{\sqrt{\lambda_3}}{\ell}
    	\overline{r}^m
        e^{-\overline{r}^2/2}
        L_n^m (\overline{r}^2)
		,
	\label{solution_R_n_m}
\end{gather}
having the normalization constant for the condition in eq. (\ref{integ_sq_many_el_WF}).

Gathering the solutions derived above, we obtain the expression for the explicitly correlated spatial WF
\begin{gather}
	\Psi_{k n m}^\pm (x_1, x_2, x_3)
    =
    	\psi^{\mathrm{HO}}_k
        \left( \frac{\sqrt{3}}{\ell} X; \frac{\ell}{\sqrt{3}} \right)
    	R_{n m} (r)
    	\Phi_m^{\pm} (\phi)
	\label{spatial_WF_3e}
\end{gather}
and its corresponding energy eigenvalue
\begin{gather}
	E^{(3)}_{k n m}
	=
        \omega_0
        \left( k + \frac{1}{2} \right)
        +
        \lambda_3
        \omega_0
	    (2 n + m + 1 )
	,
	\label{energy_eval_3}
\end{gather}
characterized by the three quantum numbers.
We should keep in mind that, at this point,
neither the Fermi statistics nor the spin states have been taken into account.

\subsubsection{Spin wave functions}

Taut et al.\cite{bib:2028} constructed approximate three-electron WFs including spin parts for an interacting three-dimensional system for the perturbative analyses of correlation effects.
The construction of eigenstates for two-spin states is straightforward,
as widely instructed in textbooks of quantum mechanics and solid-state physics.
The situation is, however, become complicated when there exist three spins in a target system.
Taut et al. provided the eigenstates of total spin of the three electrons explicitly by using the representation matrices of permutations.
We adopt their manner for the construction of the three-electron WFs in the present study.

The following four linear combinations of the $2^3 = 8$ bases for spin WFs form the spin quartet ($S = 3/2$) states:
\begin{gather}
	\chi^{3/2,3/2} (s_1, s_2, s_3)
    =
    	\alpha (s_1)
    	\alpha (s_2)
    	\alpha (s_3)
	\\
	\chi^{3/2,1/2} (s_1, s_2, s_3)
    =
    	\frac{1}{\sqrt{3}}
        [
	    	\alpha (s_1)
	    	\alpha (s_2)
	    	\beta (s_3)        	
	\nonumber \\
			+
	    	\alpha (s_1)
	    	\beta (s_2)        	
	    	\alpha (s_3)
            +
	    	\beta (s_1)        	
	    	\alpha (s_2)
	    	\alpha (s_3)
        ]
	\\
	\chi^{3/2,-1/2} (s_1, s_2, s_3)
    =
    	\frac{1}{\sqrt{3}}
        [
	    	\beta (s_1)
	    	\beta (s_2)
	    	\alpha (s_3)        	
	\nonumber \\
            +
	    	\beta (s_1)
	    	\alpha (s_2)        	
	    	\beta (s_3)
            +
	    	\alpha (s_1)        	
	    	\beta (s_2)
	    	\beta (s_3)
        ]
    \\
	\chi^{3/2,-3/2} (s_1, s_2, s_3)
    =
	    	\beta (s_1)
	    	\beta (s_2)
	    	\beta (s_3)
	.
\end{gather}
It is obvious that
these spin WFs are symmetric under the exchange of an arbitrary two spin variables.
When a spatial WF $f (x_1, x_2, x_3) \equiv f (123)$ is given,
one can easily construct the corresponding three-electron WF for $S = 3/2$ state as
\begin{gather}
	f^{3/2, S_z} (x_1, s_1, x_2, s_2, x_3, s_3)
	\nonumber \\
	=
    	\chi^{3/2, S_z} (s_1, s_2, s_3)
        \mathcal{A} f (x_1, x_2, x_3)
	\label{antisymmetrized_f_spin_32}
	,
\end{gather}
where the anti-symmetrization symbol $\mathcal{A}$ acts as
\begin{gather}
	\mathcal{A} f (123)
    \equiv
    	\sum_{\sigma}
        \mathrm{sgn} (\sigma)
        \mathcal{P}_\sigma
        f (123)
	\nonumber \\
	=        
		f (123)
        -
        f (213)
        -
        f (132)
	\nonumber \\
        -
        f (321)
        +
        f (231)
        +
        f (312)
	.        
\end{gather}
$\sigma$ is an element of the set $\{ 1, (1,2), (2,3), (1,3), (1,2,3), (1,3,2) \}$
of permutations for $\{ 1,2,3 \}$ and
$\mathcal{P}_\sigma$ is the corresponding operator for the three variables.
The WF in eq. (\ref{antisymmetrized_f_spin_32}) is clearly anti-symmetric under an arbitrary exchange of two electrons.

The remaining four spin bases form two doublets for spin singlet ($S = 1/2$) states:
\begin{gather}
	\chi^{1/2,1/2}_1 (s_1, s_2, s_3)
    =
    	\frac{1}{\sqrt{6}}
        [
    		-
	    	\alpha (s_1)        	
	    	\beta (s_2)
	    	\alpha (s_3)        	
	\nonumber \\
			-
	    	\beta (s_1)
	    	\alpha (s_2)        	
	    	\alpha (s_3)        	
            + 2
	    	\alpha (s_1)        	
	    	\alpha (s_2)        	
	    	\beta (s_3)
		]
	\label{spin_wf_doublet_1_up}        
	\\        
	\chi^{1/2,-1/2}_1 (s_1, s_2, s_3)
    =
    	\frac{1}{\sqrt{6}}
        [
	    	\beta (s_1)
	    	\alpha (s_2)        	
	    	\beta (s_3)
	\nonumber \\
            +
	    	\alpha (s_1)        	
	    	\beta (s_2)
	    	\beta (s_3)
            - 2
	    	\beta (s_1)
	    	\beta (s_2)
	    	\alpha (s_3)        	
		]
	\label{spin_wf_doublet_1_dn}        
\end{gather}
as the first one and
\begin{gather}
	\chi^{1/2,1/2}_2 (s_1, s_2, s_3)
	\nonumber \\
    =
    	\frac{1}{\sqrt{2}}
        [
	    	\alpha (s_1)        	
	    	\beta (s_2)
	    	\alpha (s_3)        	
            -
	    	\beta (s_1)
	    	\alpha (s_2)        	
	    	\alpha (s_3)        	
       	]
	\label{spin_wf_doublet_2_up}        
	\\        
	\chi^{1/2,-1/2}_2 (s_1, s_2, s_3)
	\nonumber \\
    =
    	\frac{1}{\sqrt{2}}
        [
        	-
	    	\beta (s_1)
	    	\alpha (s_2)        	
	    	\beta (s_3)
            +
	    	\alpha (s_1)        	
	    	\beta (s_2)
	    	\beta (s_3)
       	]
	\label{spin_wf_doublet_2_dn}        
\end{gather}
as the second one.\cite{bib:2028}
$\chi^{1/2, S_z}_1 (\chi^{1/2, S_z}_2)$ is symmetric (anti-symmetric) under the exchange of $s_1$ and $s_2$,
whereas it is not symmetric (anti-symmetric) under the other exchanges.
In fact, construction of a spin eigenstate of $S^2$ and $S_z$ for
three electrons that is symmetric or anti-symmetric
under an arbitrary exchange is impossible.
$\chi^{1/2, S_z}_1$ and $\chi^{1/2, S_z}_2$ for a given $S_z$ are mixed with each other when a permutation operation is applied to the spin variables.
Specifically, for a permutation $\sigma$,
the action of its corresponding operator to an $S = 1/2$ state is expressed as
\begin{gather}
	\mathcal{P}_{\sigma}
    \chi^{1/2,S_z}_j
    (s_1, s_2, s_3)
    =
    	\sum_{j'=1,2}
	    \chi^{1/2,S_z}_{j'}
	    (s_1, s_2, s_3)
        (P_\sigma)_{j' j}
\end{gather}
for $j = 1, 2$ with the representation matrices $P_\sigma$\cite{bib:2028}:
\begin{gather}
	P_{1}
    =
    	\begin{pmatrix}
        	1 & 0 \\
            0 & 1
    	\end{pmatrix}
		,
	P_{(1, 2)}
    =
    	\begin{pmatrix}
        	1 & 0 \\
            0 & -1
    	\end{pmatrix}
	,
	\nonumber \\        
	P_{(2, 3)}
    =
		\frac{1}{2}
    	\begin{pmatrix}
        	-1 & \sqrt{3} \\
            \sqrt{3} & 1
    	\end{pmatrix}
	, \,
	P_{(1, 3)}
    =
		\frac{1}{2}
    	\begin{pmatrix}
        	-1 & -\sqrt{3} \\
            -\sqrt{3} & 1
    	\end{pmatrix}
        ,
	\nonumber \\        
	P_{(1, 2, 3)}
    =
		\frac{1}{2}
    	\begin{pmatrix}
        	-1 & \sqrt{3} \\
            -\sqrt{3} & -1
    	\end{pmatrix}
	, \,        
	P_{(1, 3, 2)}
    =
		\frac{1}{2}
    	\begin{pmatrix}
        	-1 & -\sqrt{3} \\
            \sqrt{3} & -1
    	\end{pmatrix}
	.
	\label{P_matrix_spin_half}        
\end{gather}
When a spatial function $f$ is given,
these matrices enable one to construct the three-electron WFs for $S = 1/2$ states as
\begin{gather}
	f^{1/2, S_z, j}
    (x_1, s_1, x_2, s_2, x_3, s_3)
    \nonumber \\
	=
		\sum_{j', \sigma}
        \chi_{j'}^{1/2, S_z} (s_1, s_2, s_3)
        \mathrm{sgn} (\sigma)
        (P_{\sigma})_{j' j}
        \mathcal{P}_\sigma
		f (x_1, x_2, x_3)
	,
	\label{antisymmetrized_f_spin_half}
\end{gather}
apart from an appropriate normalization constant.
One can confirm that the WF in eq. (\ref{antisymmetrized_f_spin_half}) is anti-symmetric under an arbitrary exchange of two electrons by using the fact that the permutations form a group.

\subsubsection{Simultaneous eigenstates of energy and spin}

We are now prepared to construct the simultaneous eigenstates of the energy,
the total spin $S$, and the $z$ component $S_z$ of total spin.

Since $\phi (123)$ is anti-symmetric under the exchange of $x_1$ and $x_2$ as stated above,
$\Phi_m^+ (\phi (123))$ vanishes when $\mathcal{A}$ is applied:
$\mathcal{A} \Phi_m^+ (\phi(123)) = 0$,
indicating that an eigenstate for $S = 3/2$ cannot be constructed from the spatial WFs in eq. (\ref{spatial_WF_3e}) containing $\Phi_m^+$.
On the other hand, the application of $\mathcal{A}$ to $\Phi_m^- (\phi (123))$ can generate a non-vanishing function:
\begin{gather}
	\mathcal{A}
    \Phi_m^- (\phi(123))
	\nonumber \\
	=
    	2 
        [
        	\Phi_m^- (\phi(123))
            +
        	\Phi_m^- (\phi(231))
            +
        	\Phi_m^- (\phi(312))
		]
    \nonumber \\
    =
		\begin{cases}
			6 \Phi_m^- (\phi (123)) & (p = 0) \\
            0 & (p = 1, 2)
		\end{cases}
	,
\end{gather}
where we express $m = 3 m' + p$ with $p = 0, 1, 2$ and we used eq. (\ref{phi123_phi231_phi312_simple}).
We can thus construct the simultaneous eigenstate of the energy, $S = 3/2$, and $S_z$ by using eq. (\ref{antisymmetrized_f_spin_32}) for $\Psi_{k n m}^-$ in eq. (\ref{spatial_WF_3e})
with $p = 0 \, (m = 3, 6, 9, \dots)$ as
\begin{gather}
	\Psi_{k n m}^{S_z}
    (x_1, s_1, x_2, s_2, x_3, s_3)
	=
    	\Psi_{k n m}^- (123)
		\chi^{3/2, S_z} (123)
		.
	\label{WF3e_spin_3_2_unnorm}        
\end{gather}

For the construction of three-electron WFs for $S = 1/2$ states,
we adopt eq. (\ref{antisymmetrized_f_spin_half}) for $\Psi_{k n m}^\pm$
by referring to the representation matrices in eq. (\ref{P_matrix_spin_half}).
The anti-symmetrized WFs from $\Psi_{k n m}^+$ for $j = 1$
and that from $\Psi_{k n m}^-$ for $j = 2$ vanish:
$
\Psi_{k n m}^{1/2, S_z, 1 +}
=
\Psi_{k n m}^{1/2, S_z, 2 -}
=
0
.
$
On the other hand,
the anti-symmetrized WF from $\Psi_{k n m}^-$ for $j = 1$
can be a non-vanishing function given by
\begin{widetext}
\begin{gather}
	\Psi_{k n m}^{1/2, S_z, 1 -}
    (x_1, s_1, x_2, s_2, x_3, s_3)
	\nonumber \\
	=
    	\frac{\chi_1^{1/2, S_z} (123)}{3 \sqrt{2}}
        [
        	2
            \Psi_{k n m}^- (123)
            -
            \Psi_{k n m}^- (231)
            -
            \Psi_{k n m}^- (312)
		]
		+
    	\frac{\chi_2^{1/2, S_z} (123)}{3 \sqrt{2}}
        [
            -
            \sqrt{3}
            \Psi_{k n m}^- (231)
            +
            \sqrt{3}
            \Psi_{k n m}^- (312)
		]
	\nonumber \\
    =
    	\begin{cases}
        	0 & (p = 0) \\
        	[
		    	\chi_1^{1/2, S_z} (123)
				\Psi_{k n m}^- (123) 
				-
	    		\chi_2^{1/2, S_z} (123)
				\Psi_{k n m}^+ (123)
            ]/\sqrt{2}
            & (p = 1)
            \\
            [
		    	\chi_1^{1/2, S_z} (123)
				\Psi_{k n m}^- (123)
				+
	    		\chi_2^{1/2, S_z} (123)
				\Psi_{k n m}^+ (123)
			]/\sqrt{2}                
            & (p = 2)
		\end{cases}
	,
\end{gather}
\end{widetext}
where we used eq. (\ref{phi123_phi231_phi312_simple}) and introduced a normalization constant.
Although the anti-symmetrized WF $\Psi_{k n m}^{1/2, S_z, 2 +}$ from $\Psi_{k n m}^+$ for $j = 2$ can be non-vanishing as well,
one can confirm that it is the same WF as $\Psi_{k n m}^{1/2, S_z, 1 -}$.
The simultaneous eigenstates of the energy, $S = 1/2$, and $S_z$ are thus constructed from
$\Psi_{k n m}^\pm$ with $p \ne 0$ as
\begin{gather}
	\Psi_{k n m}^{S_z} (x_1, s_1, x_2, s_2, x_3, s_3)
	\nonumber \\
	=
    	\Psi_{k n m}^{1/2, S_z, 1 -} (x_1, s_1, x_2, s_2, x_3, s_3)
	\label{WF3e_spin_1_2}
	.
\end{gather}

We have constructed explicitly the simultaneous eigenstates of the energy and total spin of the three-electron system,
as the first main result of the present study.
The simultaneous eigenstates for a non-interacting case are provided in Appendix \ref{appendix:non_int_3e}.

\subsubsection{Energy spectra}

\begin{figure}
\begin{center}
\includegraphics[width=7.5cm]{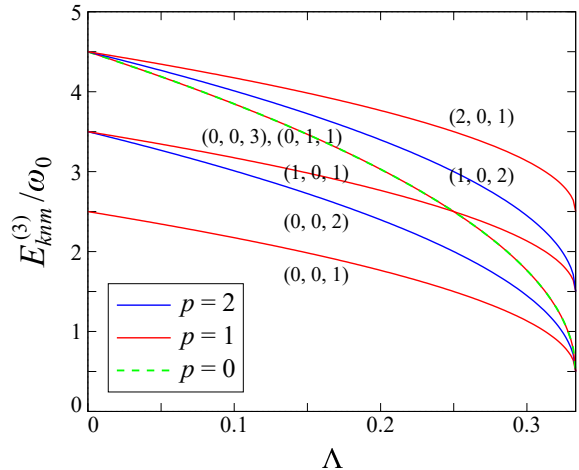}
\end{center}
\caption{
Energy eigenvalues $E^{(3)}_{k n m}$ in Table \ref{table_3e_energy} as functions of $\Lambda$.
For each of the curves, the corresponding quantum numbers $(k, n, m)$ are shown near it. 
}
\label{Fig_energy_3e}
\end{figure}

Some of the lowest eigenvalues $E^{(3)}_{k n m}$ in eq. (\ref{energy_eval_3})
and the corresponding three-electron WFs, the total spin,
and the degeneracy are shown in Table \ref{table_3e_energy}.
The ground states are doubly degenerate $S = 1/2$ states, corresponding to $(k, n, m) = (0, 0, 1)$. 
The energy eigenvalues in Table \ref{table_3e_energy} as functions of $\Lambda$ are plotted in Fig. \ref{Fig_energy_3e}.
The energy eigenvalues do not depend explicitly on the spin of the three-electron state
since the system is time-reversal invariant.

By comparing the energy eigenvalues for the interacting case in Table \ref{table_3e_energy} 
and those for the non-interacting case in Table \ref{table_non_int_energy},
we find that the four-fold degenerate non-interacting states with
$E^{(3) \mathrm{non-int}}/\omega_0 = 7/2$ split into the two doubly degenerate states when the interaction is turned on.
We find similarly that the ten-fold degenerate non-interacting states with
$E^{(3) \mathrm{non-int}}/\omega_0 = 9/2$ split into the six-fold degenerate states and the two doubly degenerate states when the interaction is turned on.
These observations corroborate our construction of the simultaneous eigenstates of the energy and total spin for the interacting system.

The electron density of the ground state for the three-electron system is now available since we know the exact WF,
it will be interesting to compare the exact energy spectra and those calculated within DFT.\cite{bib:2819,bib:4476,bib:2046,bib:4477,bib:903,bib:4478}
We did not adopt the so-called soft Coulomb interaction,
which is often used in modeling one-dimensional systems\cite{bib:3919},
since we let the analytical solvability come first.
If we resort to numerical solutions of the Schr\"odinger equation,
the soft Coulomb interaction allows us to study more realistic models and
to discuss the differences in WFs and GFs from the present model.

\begin{table}
\centering
\caption{Some of the lowest eigenvalues $E^{(3)}_{k n m}$ of the interacting three-electron system.
The three-electron wave functions, the total spin $S$, and the degeneracy are also shown.}
\label{table_3e_energy}
\begin{tabular}{cccc} 
\hline
$E^{(3)}_{k n m}/\omega_0$ & $\Psi_{k n m}^{S_z}$ & $S$ & degeneracy  \\ 
\hline
$1/2 + 2 \lambda_3$        & $\Psi_{0 0 1}^{S_z}$ & 1/2 & 2           \\
$1/2 + 3 \lambda_3$        & $\Psi_{0 0 2}^{S_z}$ & 1/2 & 2           \\
$3/2 + 2 \lambda_3$        & $\Psi_{1 0 1}^{S_z}$ & 1/2 & 2           \\
$1/2 + 4\lambda_3$        & $\Psi_{0 0 3}^{S_z}$ & 3/2 & 4           \\
\multicolumn{1}{l}{}       & $\Psi_{0 1 1}^{S_z}$ & 1/2 & 2           \\
$3/2 + 3 \lambda_3$        & $\Psi_{1 0 2}^{S_z}$ & 1/2 & 2           \\
$5/2 + 2 \lambda_3$        & $\Psi_{2 0 1}^{S_z}$ & 1/2 & 2           \\
\hline
\end{tabular}
\end{table}

\section{one-particle Green's functions}
\label{section:GF}

\subsection{Chemical potential and electron number}

\begin{figure}
\begin{center}
\includegraphics[width=7cm]{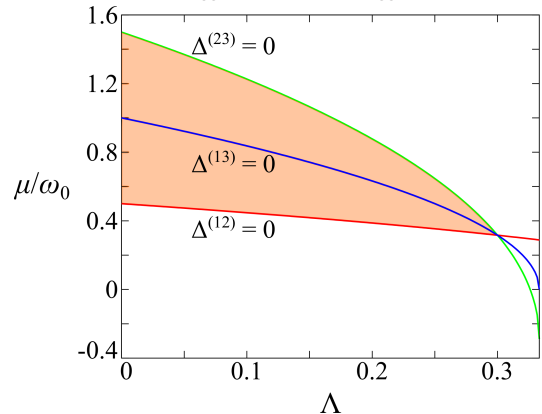}
\end{center}
\caption{
Behavior of 
$\Delta^{(12)}, \Delta^{(13)}$, and $\Delta^{(23)}$
as functions of $\Lambda$ and $\mu$.
Only when the point $(\Lambda, \mu)$ is in the shaded region,
$N_e = 2$.
The three curves cross at $(\Lambda = 0.3, \mu = \omega_0/\sqrt{10})$.
}
\label{Fig_num_el}
\end{figure}

Since we consider the one-particle GF of the two-electron system at a zero temperature ($T = 0$),
we have to take into account only the electron numbers up to 3.
The expected electron number is thus given by
$
N_e
=
Z^{-1}
\sum_{N = 1}^3
\sum_{\nu}
N
\exp [ - (E^{(N)}_\nu - \mu N)/ (k_{\mathrm{B}} T) ]
,
$
where $Z$ is the partition function and $\mu$ is the chemical potential.
$\nu$ labels a many-body energy eigenstate for a fixed $N$.
$N_e$ coincides with $N$ such that $E^{(N)}_\nu - \mu N$ is the lowest among all the possible $N$ and $\nu$ in the present case.
To examine the allowed combinations of $\Lambda$ and $\mu$ for realization of $N_e = 2$,
we define the following quantities from
eqs. (\ref{energy_eval_2}), and (\ref{energy_eval_3}):
\begin{gather}
	\Delta^{(12)}
    \equiv
		( E^{(1)}_0 - \mu)
        -
		( E^{(2)}_{0 0} - 2 \mu)
	=
    	-
        \frac{\lambda_2}{2}
        \omega_0
        +
        \mu
	\\        
	\Delta^{(13)}
    \equiv
		( E^{(1)}_0 - \mu)
        -
		( E^{(3)}_{0 0 1} - 3 \mu)
	=
    	-2
        \lambda_3
        \omega_0
        +2
        \mu
	\\
	\Delta^{(23)}
	\equiv
		( E^{(2)}_{0 0} - 2 \mu)
        -
		( E^{(3)}_{0 0 1} - 3 \mu)
	\nonumber \\
	=
        \left( \frac{\lambda_2}{2} - 2 \lambda_3 \right)
		\omega_0
		+
        \mu
\end{gather}
The behavior of $\Delta^{(12)}, \Delta^{(13)}$, and $\Delta^{(23)}$ as functions of
$\Lambda$ and $\mu$ are plotted in Fig. \ref{Fig_num_el},
indicating that $N_e = 2$ can be realized depending on the combination of the interaction strength and the chemical potential.
The largest allowed value of the interaction strength is $\Lambda_{\mathrm{max}} = 0.3$ 
for realization of $N_e = 2$, as found in the figure.

The one-, two-, and three-electron systems are equally stable at the critical point $(\Lambda = 0.3, \mu = \omega_0/\sqrt{10})$ in Fig. \ref{Fig_num_el}.
Since the degeneracies of ground states for the $N$-electron systems $n_{\mathrm{gs}}^{(N)}$ are
$n_{\mathrm{gs}}^{(1)} = 2, n_{\mathrm{gs}}^{(2)} = 1$, and $n_{\mathrm{gs}}^{(3)} = 2$,
we can calculate easily the fluctuation of electron number at the critical point as
$\delta N_e^2 = \langle N^2 \rangle - \langle N \rangle^2 = 4/5$,
which is interestingly nonzero despite the zero temperature.

We assume a combination of $\Lambda$ and $\mu$ for $N_e = 2$ in what follows.

\subsection{Interacting Green's functions}

\subsubsection{Spectral representation}

Since the ground state of the two-electron system given in eq. (\ref{WF2e_gs}) is non-degenerate,
the partition function is unity: $Z = 1$.
The one-particle GF for a complex frequency $z$ is then written as
\begin{gather}
	G_{s s'} (x, x', z)
	=
		G_{s s'}^{(\mathrm{h})} (x, x', z)
		+
		G_{s s'}^{(\mathrm{e})} (x, x', z)
	,
	\label{G_imag_freq_sum_partial_G}
\end{gather}
where
\begin{gather}
	G_{s s'}^{(\mathrm{h})} (x, x', z)
	\nonumber \\
	=
		\sum_\kappa
		\frac{
			\langle \Psi^{(2)}_{\mathrm{gs}} |
            \hat{\psi}^\dagger (x', s')
            | \Psi^{(1)}_\kappa \rangle
			\langle \Psi^{(1)}_\kappa |
            \hat{\psi} (x, s)
            | \Psi^{(2)}_{\mathrm{gs}} \rangle
		}{
			z + \mu + E^{(1)}_\kappa - E^{(2)}_{\mathrm{gs}}
		}
	\label{def_partial_G_h}
\end{gather}
and
\begin{gather}
	G_{s s'}^{(\mathrm{e})} (x, x', z)
	\nonumber \\
	=
		\sum_\kappa
		\frac{
			\langle \Psi^{(2)}_{\mathrm{gs}} |
            \hat{\psi} (x, s)
            | \Psi^{(3)}_\kappa \rangle
			\langle \Psi^{(3)}_\kappa |
            \hat{\psi}^\dagger (x', s')
            | \Psi^{(2)}_{\mathrm{gs}} \rangle
		}{
			z + \mu + E^{(2)}_{\mathrm{gs}} - E^{(3)}_\kappa
		}
	\label{def_partial_G_e}
\end{gather}
are the partial GFs\cite{bib:3066} at a zero temperature
for hole and electron excitations, respectively, in spectral representation.
$\kappa$ in eqs. (\ref{def_partial_G_h}) and (\ref{def_partial_G_e}) runs over all the many-body energy eigenstates for a given electron number.

\begin{widetext}

\subsubsection{Hole excitation}

For the spatial parts of the one- and two-electron WFs, 
we define the integral
\begin{gather}
	M_n (x_2)
	\equiv
	\int_{-\infty}^\infty
	d x_1
		\psi_n^{\mathrm{HO}} \left( \frac{x_1}{\ell}; \ell \right)^*
		\Psi_{00} ( x_1, x_2)
	\nonumber \\
    =
		\left( \frac{ m_e \omega_0 \lambda_2}{\pi^3} \right)^{1/4}
        \frac{1}{ \sqrt{2^n n!}}
		\exp
        \left[
        	-\frac{1 + 3 \lambda_2}{4 (3 + \lambda_2)} \overline{x}_2^2
		\right]
		\int_{-\infty}^\infty
        d \overline{x}_1
            H_n \left( \frac{\overline{x}_1}{\sqrt{2}} \right)
            \exp
            \left[
            	-\frac{3 + \lambda_2}{8}
                \left(
                	\overline{x}_1
                    +
                    \frac{1 - \lambda_2}{3 + \lambda_2}
                    \overline{x}_2
                \right)^2
            \right]
	,            
	\label{matel_psi_n_Psi_00_spatial}
\end{gather}
where $\overline{x}_2 \equiv x_2/\ell$
and eq. (\ref{harm_delta_two_el_wave_func}) was used.
With this, the matrix element of the annihilation operator [see eq. (\ref{matel_creation_opr_as_integ_of_WFs})],
often called the quasihole wave function,
is written as
\begin{gather}
	\langle \Psi_n^{S_z} |
    \hat{\psi} (x_2, s_2)
    | \Psi_{\mathrm{gs}}^{(2)} \rangle
	=
    	\sum_{s_1}
        \int_{-\infty}^\infty
        d x_1
			\Psi_n^{S_z} ( x_1, s_1 )^*
            \Psi_{\mathrm{gs}}^{(2)} ( x_1, s_1, x_2, s_2 )        
    =
        \begin{cases}
        	  M (x_2) \beta (s_2)/\sqrt{2} & (S_z =  1/2) \\
        	- M (x_2 )\alpha (s_2)/\sqrt{2} & (S_z = -1/2) \\
        \end{cases}
	.
	\label{matel_psi_n_Psi_00}
\end{gather}
Substitution of this into eq. (\ref{def_partial_G_h})
and the completeness of spin WFs,
$
\alpha (s)
\alpha (s')^*
+
\beta (s)
\beta (s')^*
=
\delta_{s s'}
,
$
lead to
\begin{gather}
	G_{s s'}^{(\mathrm{h})} (x, x', z)
	\nonumber \\
	=
		\sum_{n = 0}^\infty
        \sum_{S_z}
		\frac{
			\langle \Psi^{(2)}_{\mathrm{gs}} |
            \hat{\psi}^\dagger (x', s')
            | \Psi_n^{S_z} \rangle
			\langle \Psi_n^{S_z} |
            \hat{\psi} (x, s)
            | \Psi^{(2)}_{\mathrm{gs}} \rangle
		}{
			z + \mu + (n - \lambda_2/2) \omega_0 
		}
    =
    	\delta_{s s'}
    	\frac{1}{2}
		\sum_{n = 0}^\infty
		\frac{
        	M_n (x')^*
	        M_n (x)
		}{z + \mu + (n - \lambda_2/2) \omega_0 }
    =
    	\delta_{s s'}
		G^{(\mathrm{h})} (x, x', z)
	,
	\label{partial_GF_hole}        
\end{gather}
where the spin-independent GF
\begin{gather}
	G^{(\mathrm{h})} (x, x', z)
    =
    	\frac{1}{2}
		\sqrt{ \frac{ m_e \lambda_2}{\pi^3 \omega_0} }
		\Gamma (\nu)
        F^{(\mathrm{h})}
        \left( \frac{x}{\ell}, \frac{x'}{\ell}, z \right)
        \Bigg|_{\nu = (z + \mu)/\omega_0 - \lambda_2/2}      
	\label{G_h_spin_indep}
\end{gather}
is calculated from the dimensionless quantity
\begin{gather}
	F^{(\mathrm{h})} (\overline{x}, \overline{x}', z)
    =
    	\int_{-\infty}^\infty
        d \zeta
        	K^+ (\zeta, \overline{x}, z)
    	\int_{-\infty}^{\zeta}
        d \zeta'
        	K^- (\zeta', \overline{x}', z)
    	+
    	\int_{-\infty}^\infty
        d \zeta
        	K^- (\zeta, \overline{x}, z)
    	\int_{\zeta}^\infty
        d \zeta'
        	K^+ (\zeta', \overline{x}', z)
\end{gather}
with
\begin{gather}
	K^\pm (\zeta, \overline{x}, z)
	\equiv
			\exp 
            \left(
				-
                \frac{1 + \lambda_2}{8}
               	\zeta^2
                -
                \frac{1 - \lambda_2}{4}
                \zeta
				\overline{x}
                -
                \frac{1 + \lambda_2}{8}
				\overline{x}^2
			\right)
            D_{-\nu} ( \pm \zeta )
	.
    \label{def_K_for_GF_hole}
\end{gather}
\end{widetext}
To get the expression in eq. (\ref{def_K_for_GF_hole}),
we used the formula provided by Glasser and Nieto\cite{bib:3963},
which expresses the infinite summation over the Hermite polynomials via the parabolic cylinder function $D_\nu$\cite{bib:3956}.

The pole positions of $G^{(\mathrm{h})}$ on $z$ axis are those of the Gamma function in eq. (\ref{G_h_spin_indep}),
which are given by
\begin{gather}
	\frac{z + \mu}{\omega_0}
    -
    \frac{\lambda_2}{2}
    =
    	0, -1, -2, \dots
	\label{pole_pos_GF_hole}
\end{gather}
Each of these pole positions corresponds to a certain $n$ in eq. (\ref{partial_GF_hole}),
indicating that the pole for $n = 0$ represents a one-electron system
where the electron occupies the lowest-energy orbital after the removal of an electron from the two-electron system.
This process contributes as a major peak of the imaginary part of $G^{(\mathrm{h})}$ on $z$ axis,
known as a quasiparticle peak.
The other processes ($n \ne 0$) represent the systems,
in each of which the remaining electron is excited.
They contribute as minor peaks known as satellite peaks.
The obvious difference in the strength between these two kinds of peaks for hole excitations will be demonstrated later.

\subsubsection{Electron excitation}

Since the ground state of the two-electron system is of $S = 0$,
no transition to an $S = 3/2$ state occurs when an electron is added to the system: 
\begin{gather}
	\langle \Psi_{k n m}^{S_z} |
    \hat{\psi}^\dagger (x, s)
    | \Psi_{\mathrm{gs}}^{(2)} \rangle
    =
    	0
        \, (p = 0)
	.
\end{gather}

By using the integral
\begin{gather}
    M_{k n m}^\pm (x_3)
    \equiv
		\int
	    d x_1
		d x_2
    		\Psi_{k n m}^\pm (x_1, x_2, x_3)^*
        	\Psi_{00} (x_1, x_2)
	\label{def_matel_spatial_x3}
\end{gather}
for the spatial parts of the two- and three-electron WFs with $p = 1, 2$ and 
\begin{gather}
	M_j^{S_z} (s_3)
    \equiv
		\sum_{s_1, s_2}
			\chi_j^{1/2, S_z} (s_1, s_2, s_3)^*
			\chi^{0, 0} (s_1, s_2)
	\label{def_matel_spin_s3}
\end{gather}
for the spin parts (see Appendix \ref{appendix:calc_M_spin}), 
the matrix element of the creation operator [see eq. (\ref{matel_creation_opr_as_integ_of_WFs})],
often called the quasiparticle wave function,
is written as
\begin{widetext}
\begin{gather}
	\langle \Psi_{k n m}^{S_z} |
    \hat{\psi}^\dagger (x_3, s_3)
    | \Psi_{\mathrm{gs}}^{(2)} \rangle
	\nonumber \\
	=
   	\frac{1}{2 !}
	\sum_{s_1, s_2}
	\int
    d x_1
    d x_2
    	\Psi_{k n m}^{S_z} (x_1, s_1, x_2, s_2, x_3, s_3)^*
        \Psi_{\mathrm{gs}}^{(2)} (x_1, s_1, x_2, s_2)
    =
		\mp
    	\frac{1}{2 \sqrt{2}}
    	M_2^{S_z} (s_3)
		M_{k n m}^+ (x_3)
	,
	\label{matel_creation_WF3_WF2gs}
\end{gather}
where the negative (positive) sign on the right-hand side is for $p = 1 \, (p = 2)$.
The explicit expression of $M_{k n m}^+ (x_3)$ is derived in Appendix \ref{appendix:calc_M_spatial}.
Substitution of eq. (\ref{matel_creation_WF3_WF2gs}) into eq. (\ref{def_partial_G_e})
and the completeness of spin WFs lead to
\begin{gather}
	G_{s s'}^{(\mathrm{e})} (x, x', z)
	=
    	\delta_{s s'}
    	\frac{1}{8}
		\sum_{k, n, m (p = 1, 2)}^\infty
		\frac{
    	    M_{k n m}^+ (x)^*
        	M_{k n m}^+ (x')
		}{
			z + \mu
            -
            ( k + d_{n m} )
	        \omega_0
		}
	=
    	\delta_{s s'}
		\sum_{n, m (p = 1, 2)}^\infty
			G_{n m}^{(\mathrm{e})} (x, x', z)
	,
	\label{partial_G_e_in_tilde_M}
\end{gather}
where $d_{n m} \equiv \lambda_3 (2 n + m + 1) - \lambda_2/2$
and the spin-independent GF
\begin{gather}
	G_{n m}^{(\mathrm{e})} (x, x', z)
    =
    	-
        4
        \sqrt{\frac{m_e \lambda_2}{\pi^5 \omega_0}}
            \frac{n !}{(n + m)!}
	        \Gamma (\nu_{n m})
			F^{(\mathrm{e})}_{n m}
            \left(
            	\sqrt{3} \frac{x}{\ell},
                \sqrt{3} \frac{x'}{\ell},
                z
            \right)
            \Bigg|_{\nu_{n m} = - (z + \mu)/\omega_0 + d_{n m} }
	\label{G_n_m_e_spin_indep}
\end{gather}
is calculated from the dimensionless quantity
\begin{gather}
	F^{(\mathrm{e})}_{n m} (\overline{x}, \overline{x}', z)
	\nonumber \\
	=
		\int_{-\infty}^\infty
		d \zeta
        	K_{n m}^+ (\zeta, \overline{x}, z)
    	\int_{-\infty}^{\zeta + 3 (\overline{x} - \overline{x}')/2}
        d \zeta'
        	K_{n m}^- (\zeta', \overline{x}', z)
	    +
		\int_{-\infty}^\infty
		d \zeta
        	K_{n m}^- (\zeta, \overline{x}, z)
    	\int_{\zeta + 3 (\overline{x} - \overline{x}')/2}^\infty
		d \zeta'
        	K_{n m}^+ (\zeta', \overline{x}', z)
\end{gather}
with
\begin{gather}
	K_{n m}^\pm (\zeta, \overline{x}, z)
	\equiv
        \exp
        \left[
        	-\frac{ ( \zeta + \overline{x} )^2 }{6}
		\right]
        D_{-\nu_{n m}}
        \left(
        	\pm \left\{ \frac{2}{3} \zeta + \overline{x} \right\}
		\right)
		I_{n m}
    	\left(
    		\frac{\sqrt{\lambda_3}}{3}
        	\zeta
		\right)
	,
	\label{def_K_for_GF_el}
\end{gather}
\end{widetext}
for which $I_{n m}$ is defined in eq. (\ref{def_integ_I_n_m_r0}).
To get the expression in eq. (\ref{def_K_for_GF_el}),
we used the formula\cite{bib:3963} for the infinite summation over the Hermite polynomials.

The pole positions of $G_{n m}^{(\mathrm{e})}$ on $z$ axis are those of the Gamma function in eq. (\ref{G_n_m_e_spin_indep}),
which are given by
\begin{gather}
	\frac{z + \mu}{\omega_0}
    -
    d_{n m}
    =
    	0, 1, 2, \dots
	\label{pole_pos_GF_el}
\end{gather}
Each of these pole positions corresponds to a certain $k$ in eq. (\ref{partial_G_e_in_tilde_M}) for given $n$ and $m$,
indicating that a pole for $k = 0$ $(k \ne 0)$ represents a three-electron system
where the center-of-mass motion is not excited (is excited) after the addition of an electron to the two-electron system.

We have obtained finally the exact expressions of the GF for the two-electron system,
as the second main result of the present study.
The expressions for the partial GFs we derived are more favorable 
than the generic expressions in eqs. (\ref{def_partial_G_h}) and (\ref{def_partial_G_e}) for practical calculations.
It is because the summation over one of the three quantum numbers has been already taken exactly in our expressions.

\subsubsection{Pole strengths}

\begin{figure}
\begin{center}
\includegraphics[width=8.5cm]{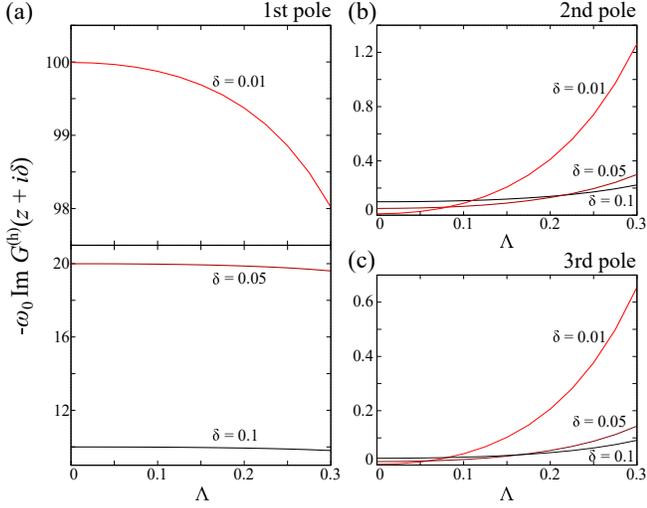}
\end{center}
\caption{
Imaginary part of integrated partial GF $G^{(\mathrm{h})}$ near its three poles closest to the origin of $z$ as function of the interaction strength.
$\delta$ is a small deviation of $z$ from the poles on the real axis.
}
\label{Fig_hole_pole_str}
\end{figure}

Photoemission spectra\cite{bib:4070,bib:4165,bib:pw_unfolding} are related to the one-particle GF integrated over spatial variable in general.
Here we examine the integrated GF of our system
\begin{gather}
	G (z)
    =
    	\int_{-\infty}^\infty
        d x \,
			G (x, x, z)
	,
\end{gather}
for which we plot the imaginary part of $G^{(\mathrm{h})} (z)$ near the poles in Fig. \ref{Fig_hole_pole_str}.
It is seen that the magnitudes near the first pole,
which is identified as the quatiparticle peak at $z + \mu = \lambda_2/2$ [see eq. (\ref{pole_pos_GF_el})],
is much larger than those near the other poles, identified as the satellite peaks.
These observations indicate that the independent-particle picture is basically valid for the hole excitation process,
realizing the quasiparticle peak,
while the many-electron effects cause the satellite peaks.

\subsection{Non-interacting Green's functions}

It is instructive to compare the interacting GFs and the corresponding (non-interacting) Kohn--Sham GFs.
We derive the expressions for the latter ones here.

\subsubsection{Hole excitation}

When the interaction is absent ($\Lambda = 0$),
the integral $M_n (x_2)$ in eq. (\ref{matel_psi_n_Psi_00_spatial}) is nonzero only for $n = 0$.
The spin-independent partial GF for hole excitations are then calculated as
\begin{gather}
	G^{(\mathrm{h}) \mathrm{non-int} } (x, x', z)
	=
		\sqrt{ \frac{ m_e }{\pi \omega_0} }
		\frac{\exp [ -(\overline{x}^2 + \overline{x}'^2)/4 ]}
        {(z + \mu)/\omega_0 - 1/2}
	,
	\label{partial_GF_hole_non_int}
\end{gather}
where $\overline{x} \equiv x/\ell$ and $\overline{x}' \equiv x'/\ell$,
as a special case of eq. (\ref{G_h_spin_indep}).

\subsubsection{Electron excitation}

For the derivation of expressions for the non-interacting GF for electron excitations,
it is easier to work with the quantum numbers $(n_1, n_2, n_3)$
for the independent oscillators described
in Appendix \ref{appendix:non_int_3e}
than those for the interacting case, $(k, n, m)$.

Since the ground state of the non-interacting two-electron system
consists of two electrons occupying the lowest-energy spatial orbital to form an $S = 0$ state,
the normalized WF is given by
$
	\Psi_{\mathrm{gs}}^{(2) \mathrm{non-int}} (x_1, x_2, s_1, s_2)
	\nonumber \\
	=
    	\sqrt{2}
		\psi_0 (x_1) \psi_0 (x_2)
        \chi^{0,0} (s_1, s_2)
	.
$
When an electron is added to the system,
the electron occupies another spatial orbital $\psi_n$ with $n \ne 0$
to form a three-electron system whose WF is given by
$\psi_{n 0 0}^{1/2, S_z}$ in eq. (\ref{eigen_func_non_int_n1_n2_n2}).
The matrix element of the creation operator is then calculated as (see Appendix \ref{appendix:calc_M_spin})
\begin{gather}
	\langle \psi_{n 0 0}^{1/2, S_z} |
	\hat{\psi}^\dagger (x_3, s_3)
	| \Psi^{(2) \mathrm{non-int}}_{\mathrm{gs}} \rangle
	\nonumber \\
	=
		\psi_n (x_3)^*
        \cdot
        \begin{cases}
        	\alpha (s_3)^* & (S_z = 1/2) \\
            -\beta (s_3)^* & (S_z = -1/2) \\
        \end{cases}
	,        
\end{gather}
which is then substituted into eq. (\ref{def_partial_G_e}) to get the spin-independent partial GF for electron excitations:
\begin{gather}
	G^{(\mathrm{e}) \mathrm{non-int}} (x, x', z)
    =
		\sum_{n = 1}^\infty
		\frac{\psi_n (x) \psi_n (x')^*}{z + \mu - (n + 1/2) \omega_0}
	\nonumber \\
	=
        \sqrt{\frac{m_e}{\pi \omega_0}}
	    \Bigg[
        	-
            \frac{\exp [ -(\overline{x}^2 + \overline{x}'^2)/4 ] }
            {(z + \mu)/\omega_0 - 1/2}
	\nonumber \\
			-
            \Gamma (\nu)
            D_{-\nu} ( \overline{x}_>)
            D_{-\nu} (-\overline{x}_<)
		\Bigg]_{\nu = - (z + \mu)/\omega_0 + 1/2}
	.        
	\label{partial_GF_el_non_int}
\end{gather}
$\overline{x}_>$ and $\overline{x}_<$ are defined for
$\overline{x}$ and $\overline{x}'$ so that
$\overline{x}_> > \overline{x}_<$.
To get the last equality in eq. (\ref{partial_GF_el_non_int}),
we used the formula for the infinite summation over the Hermite polynomials provided by Glasser and Nieto\cite{bib:3963},
who derived the expression of the non-interacting GF,
$
G^{\mathrm{non-int}}
=
G^{(\mathrm{h}) \mathrm{non-int}}
+
G^{(\mathrm{e}) \mathrm{non-int}}
$
.

\subsubsection{Kohn--Sham Green's function}

It is well known that the exact KS potential
for ground state of an interacting two-electron system can be constructed
if the electron density $n_{\mathrm{gs}}^{(2)} (\boldsymbol{r})$
of the spin-singlet ground state is known\cite{bib:3919, bib:3915, bib:3914}:
\begin{gather}
	V_{\mathrm{KS}} (\boldsymbol{r})
	=
		\frac{\nabla^2 \sqrt{n_{\mathrm{gs}}^{(2)} (\boldsymbol{r}) } }
        {2 m_e \sqrt{n_{\mathrm{gs}}^{(2)} (\boldsymbol{r}) }}
	\label{exact_V_KS_for_two_electrons}
\end{gather}

The exact KS potential for our interacting two-electron system is calculated from eqs. (\ref{WF2e_gs}) and (\ref{exact_V_KS_for_two_electrons}) as\cite{bib:4276}
\begin{gather}
	V_{\mathrm{KS}} (x)
	=
		\omega_0
		\left( \frac{\lambda_2}{1 + \lambda_2} \right)^2
		\overline{x}^2
	.        
	\label{harm_harm_V_KS}
\end{gather}
It is clear that this KS Hamiltonian for the effective non-interacting system is obtained
apart from a constant simply by replacing $\omega_0$ with
$
2 \omega_0 \lambda_2 / (1 + \lambda_2)
\equiv
\alpha_{\mathrm{KS}}^2 \omega_0
$
and setting $\Lambda = 0$ in the original Hamiltonian in eq. (\ref{def_H2}).
The $n$-th KS orbital energy is thus given by 
\begin{gather}
	\varepsilon^{\mathrm{KS}}_n
    =
    	\alpha_{\mathrm{KS}}^2 \omega_0
        \left( n + \frac{1}{2} \right)
	.
\end{gather}
With the positive constant $\alpha_{\mathrm{KS}}$,
the partial GFs of the KS system are obtained via the replacement for eqs. (\ref{partial_GF_hole_non_int}) and (\ref{partial_GF_el_non_int}) as
\begin{gather}
	G^{(\mathrm{h}) \mathrm{KS} } (x, x', z)
    =
    	-
		\sqrt{ \frac{ m_e }{\pi \omega_0} }
		\frac{1}{\alpha_{\mathrm{KS}}}
		\frac{\exp [ 
                	- \alpha_{\mathrm{KS}}^2
                    (\overline{x}^2 + \overline{x}'^2)/4 ]
		}{\nu_{\mathrm{KS}}}
\end{gather}
and
\begin{gather}
	G^{(\mathrm{e}) \mathrm{KS}} (x, x', z)
    =
        \sqrt{\frac{m_e}{\pi \omega_0}}
		\frac{1}{\alpha_{\mathrm{KS}}}
	    \Bigg[
            \frac{
            	\exp [ 
                	- \alpha_{\mathrm{KS}}^2
                    (\overline{x}^2 + \overline{x}'^2)/4 ]
			}{\nu_{\mathrm{KS}}}
	\nonumber \\
			-
            \Gamma (\nu_{\mathrm{KS}})
            D_{-\nu_{\mathrm{KS}}} ( \alpha_{\mathrm{KS}} \overline{x}_>)
            D_{-\nu_{\mathrm{KS}}} (-\alpha_{\mathrm{KS}} \overline{x}_<)
		\Bigg]
	,        
	\label{partial_GF_KS_non_int}
\end{gather}
where
$
\nu_{\mathrm{KS}}
\equiv
- (z + \mu)/(\alpha_{\mathrm{KS}}^2 \omega_0) + 1/2
.
$
We have adopted the chemical potential of the original system as that for the KS system.

\begin{figure}
\begin{center}
\includegraphics[width=7cm]{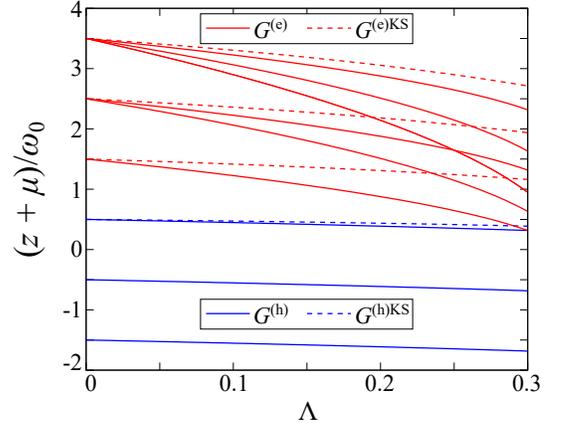}
\end{center}
\caption{
Pole positions of the interacting GF of the two-electron system as functions of the interaction strength $\Lambda$.
Those of the KS GF for each $\Lambda$ are also plotted.
For the pole positions coming from $G^{(\mathrm{e})}$ and $G^{(\mathrm{e}) \mathrm{KS}}$,
only those corresponding to the energy eigenvalues shown in Table \ref{table_3e_energy} are plotted.
}
\label{Fig_poles}
\end{figure}

The pole positions of $G$ and those of $G^{\mathrm{KS}}$ on the energy axis (real axis of $z$) as functions of $\Lambda$ are plotted in Fig. \ref{Fig_poles}.
One finds that there exist more non-degenerate poles for an interacting ($\Lambda \ne 0$) case than in the corresponding KS system.
Furthermore, the discrepancies between the pole positions of $G$ and those of $G^{\mathrm{KS}}$ become larger as the interaction becomes stronger.
There exists only one pole coming from
$G^{(\mathrm{h}) \mathrm{KS}}$,
while an infinite number of poles from $G^{(\mathrm{h})}$ exist in the negative direction of $z$ axis.
For both of $G^{(\mathrm{e})}$ and
$G^{(\mathrm{e}) \mathrm{KS}}$, on the other hand,
there exist an infinite number of poles in the positive $z$ direction.

The fundamental gap is calculated as
$
E_{\mathrm{gap}}
=
E_{\mathrm{gs}}^{(3)}
-
2
E_{\mathrm{gs}}^{(2)}
+
E_{\mathrm{gs}}^{(1)}
=
\omega_0
(2 \lambda_3 - \lambda_2)
,
$
while the HOMO-LUMO gap of the KS system is
$
E_{\mathrm{gap}}^{\mathrm{KS}}
=
\varepsilon^{\mathrm{KS}}_1
-
\varepsilon^{\mathrm{KS}}_0
=
2 \omega_0 \lambda_2  /(1 + \lambda_2)
.
$
One can easily confirm that
$E_{\mathrm{gap}} < E_{\mathrm{gap}}^{\mathrm{KS}}$
for the allowed range of $\Lambda$.
This fact is also understood by looking at Fig. \ref{Fig_poles},
where $E_{\mathrm{gap}}$ ($E_{\mathrm{gap}}^{\mathrm{KS}}$)
is nothing but the distance between the neighboring poles
on the $z$ axis coming from
$G^{(\mathrm{h})}$ and $G^{(\mathrm{e})}$
($G^{(\mathrm{h}) \mathrm{KS}}$
and
$G^{(\mathrm{e}) \mathrm{KS}}$).
These observations indicate that our system is an example where even the exact KS potential does not give the correct energy gap particularly for strong interactions.

We can also calculate the pole positions of the exact GF of the three-electron system for hole excitation.
[See Appendix \ref{appendix:hole_excitation_3e}]

\subsection{Numerical examination}

\begin{figure}
\begin{center}
\includegraphics[width=7cm]{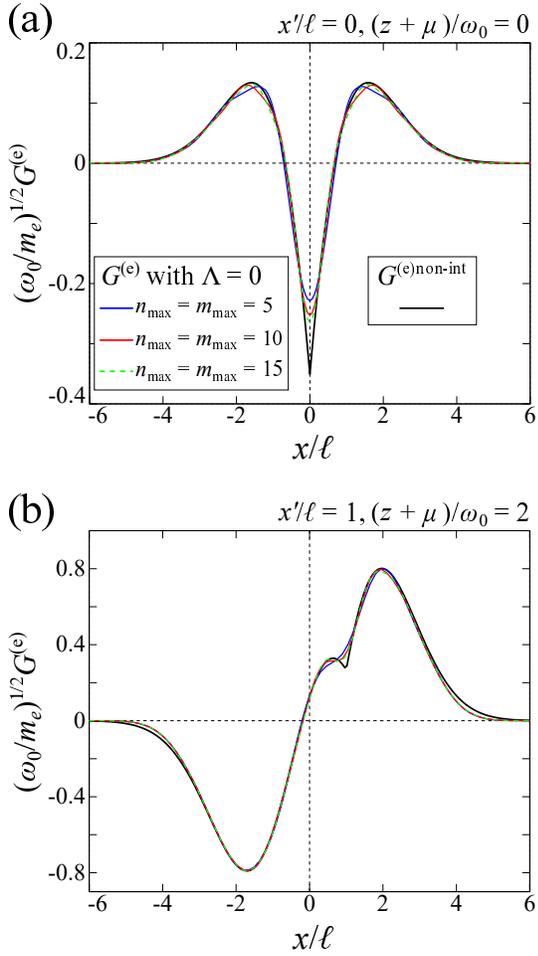}
\end{center}
\caption{
Spin independent partial GFs for electron excitation.
$G^{(\mathrm{e})}$ with $\Lambda = 0$ using the expression for the interacting GF and $G^{(\mathrm{e}) \mathrm{non-int}}$ as functions of $x$ are  plotted for
(a)
$x'/\ell = 0, (z + \mu)/\omega_0 = 0$
and 
(b)
$x'/\ell = 1, (z + \mu)/\omega_0 = 2$
.
For the expression of $G^{(\mathrm{e})}$, the GF for different truncations in infinite summations are plotted.
}
\label{Fig_gf_elec_n_m_max}
\end{figure}

Our expression of the interacting GF for electron excitations in eq. (\ref{partial_G_e_in_tilde_M}) involves the infinite summation over the two quantum numbers.
We have to truncate the summation at some values $n_{\mathrm{max}}$ and $m_{\mathrm{max}}$ for a practical calculation.
We first examined the convergence of $G^{(\mathrm{e})}$ with respect to $n_{\mathrm{max}}$ and $m_{\mathrm{max}}$.
$G^{(\mathrm{e})}$ with $\Lambda = 0$ using eq. (\ref{partial_G_e_in_tilde_M}) and
$G^{(\mathrm{e}) \mathrm{non-int}}$ using eq. (\ref{partial_GF_el_non_int}) as functions of $x$ are  plotted in Fig. \ref{Fig_gf_elec_n_m_max}.
It is seen that $G^{(\mathrm{e}) \mathrm{non-int}}$ has cusps at $x = x'$.
These features are not seen in the calculated $G^{(\mathrm{e})}$
since, in general, the summation over a finite number of smooth functions gives another smooth function.
It is also found, however, that the calculated $G^{(\mathrm{e})}$ for larger $n_{\mathrm{max}}$ and $m_{\mathrm{max}}$ has more similar shapes to $G^{(\mathrm{e}) \mathrm{non-int}}$.
The convergence is slower for $x$ near the cusps than for that away the cusps.
We adopt $n_{\mathrm{max}} = m_{\mathrm{max}} = 10$ in what follows because this truncation leads to the satisfactory convergence for qualitative discussion in the present study.

\begin{figure*}
\begin{center}
\includegraphics[width=17cm]{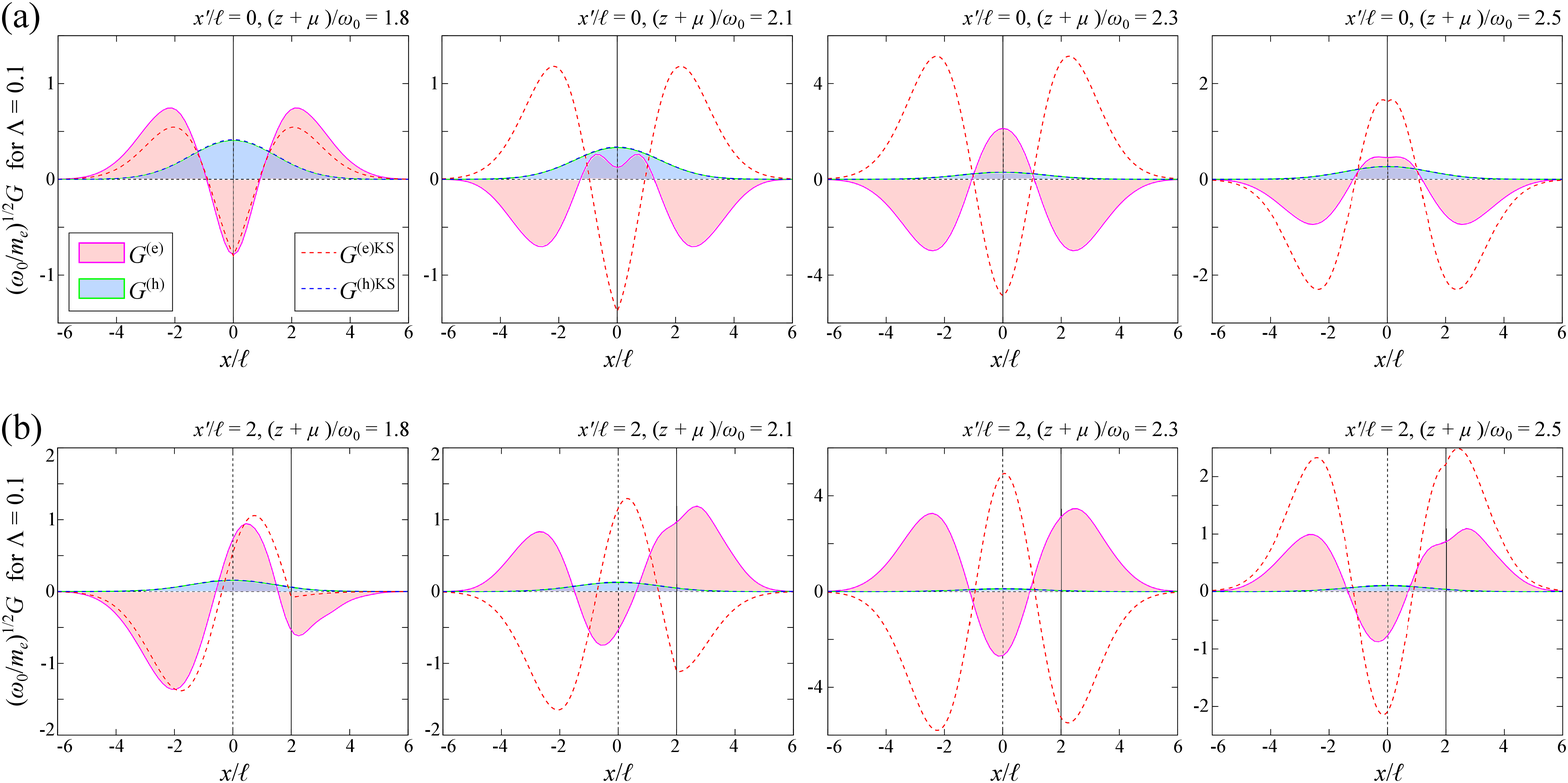}
\end{center}
\caption{
Spin independent exact partial GFs $G^{(\mathrm{e})}$ and $G^{(\mathrm{h})}$ for
$\Lambda = 0.1$ as functions of $x$ with fixed $x'$ and $z$.
The plots are for (a) $x'/\ell = 0$ and for (b) $x'/\ell = 2$.
The KS partial GFs $G^{\mathrm{(e) KS}}$ and $G^{\mathrm{(h) KS}}$ corresponding to the exact GFs are also shown.
The vertical solid lines represent the locations of cusps of $G^{\mathrm{(e) KS}}$.
}
\label{Fig_gf_x2_omg_fixed_L01}
\end{figure*}

\begin{figure*}
\begin{center}
\includegraphics[width=17cm]{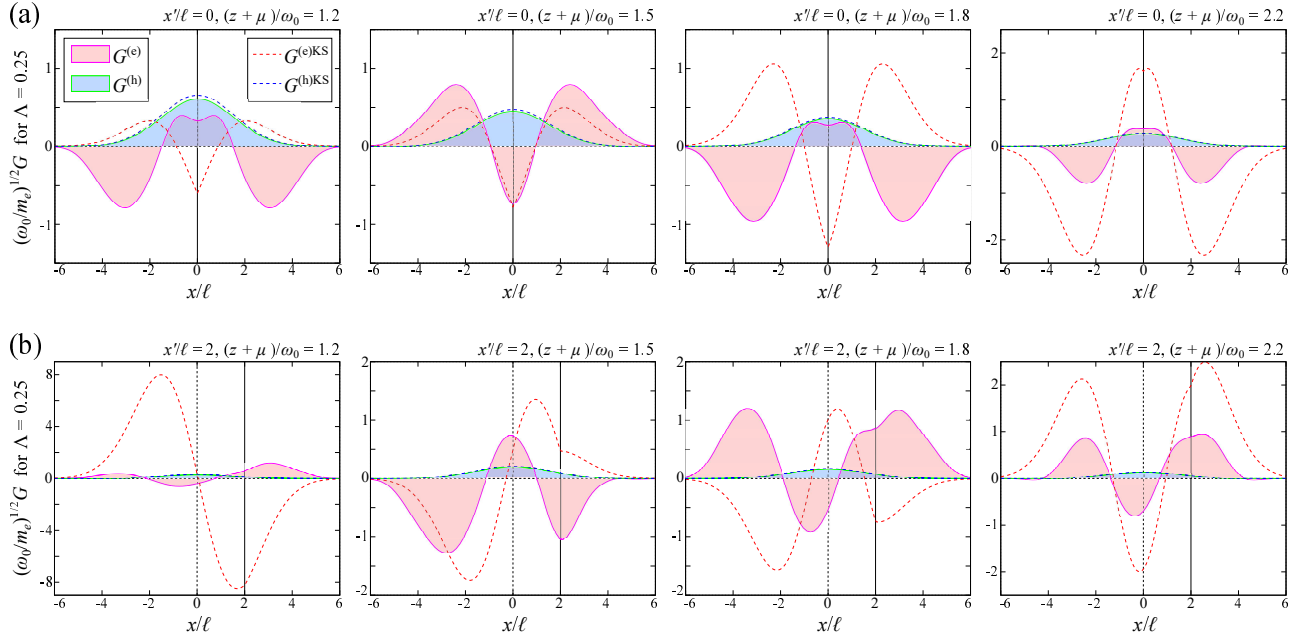}
\end{center}
\caption{
Spin independent exact partial GFs for $\Lambda = 0.25$ and their corresponding KS GFs similar to Fig. \ref{Fig_gf_x2_omg_fixed_L01}.
}
\label{Fig_gf_x2_omg_fixed_L025}
\end{figure*}

The spin independent partial GFs as functions of $x$ with fixed $x'$ and $z$ for $\Lambda = 0.1$ and $0.25$ are plotted
in Figs. \ref{Fig_gf_x2_omg_fixed_L01} and \ref{Fig_gf_x2_omg_fixed_L025}, respectively.
It is seen that the contributions from hole excitations are small compared to those from electron excitations both in the exact and the KS GFs.
The shapes of the exact and KS GFs can change drastically as $z$ crosses the poles on the frequency axis [see Fig. \ref{Fig_poles}].
More importantly, one finds that the shape and/or the magnitude of $G^{\mathrm{(e)}}$ for a given combination of $x'$ and $z$ are not similar to those of the corresponding $G^{\mathrm{(e) KS}}$ at all, depending on $x'$ and $z$.
This observation reminds us of the fact that the foundation of DFT ensures only the correctness of the total energy and the electron density for the ground state of a given electronic system.
In spite of this fact,
the quantitative predictions provided by GF-based approach in electronic-structure calculations have been successful by and large.
It is at least partially because the physical quantities to be predicted are calculated only via integration of the interacting GF of a realistic target system.
The examination on the GFs of our model system provides us a lesson that we have to be careful if we want to know the shape of the GF itself using the DFT results as a reference state.

\section{conclusions}
\label{section:conclusions}

For a three-electron system with finite-strength interactions confined to a one-dimensional harmonic trap,
we solved the Schr\"odinger equation analytically to obtain the exact solutions,
from which we constructed explicitly the simultaneous eigenstates of the energy and
the total spin for the first time.
The solutions for the three-electron system allowed us to derive the analytic expressions for the exact one-particle GF for the corresponding interacting two-electron system.
We also derived the KS GF by using the exact KS potential for the two-electron system.
We identified the pole positions of the exact and KS GFs
and found that the discrepancies between them becomes larger for a stronger interaction.

We performed numerical examination on the behavior of GFs in real space.
It was demonstrated that the exact and KS GFs can have shapes quite different from each other depending on the frequency 
and they can change drastically when the frequency crosses the poles,
implying that careful analyses are required when a GF itself is examined within a mean-field-based approach.

We have now the exact solutions for the three-electron system at hand,
it is possible to examine the many-body effects in the three-electron system without resorting to second-quantized form.
It will be interesting to look for many-body effects in the three-electron system which are absent in the two-electron system.
In addition, the two-electron system we considered will serve as a minimal real-space model having a non-trivial interacting GF.
We expect that this simple model will help to find and understand generic characteristics of GFs.

Our prescription for the construction of eigenstates for three electrons is applicable 
for three bosons by paying attention to the Bose statistics.
It is thus expected that the three-boson system equipped with the Schr\"odinger equation in the same form as in the present study will serve as a new analytically solvable model of interacting three bosons.\cite{bib:bosons1,bib:bosons2,bib:bosons3}
It is worth examining the exact GF of an interacting two-boson system.

\begin{acknowledgments}

This research was supported by MEXT as Exploratory Challenge on Post-K computer (Frontiers of Basic Science: Challenging the Limits).
This research used computational resources of the K computer provided by the RIKEN Advanced Institute for Computational Science through the HPCI System Research project (Project ID: hp180227).

\end{acknowledgments}

\appendix

\section{many-electron wave functions}

The normalization conditions and the definitions of matrix elements involving many-electron WFs are the same as in Ref. \cite{stefanucci2013nonequilibrium}.
In order for this paper to be self-contained,
we summarize the expressions necessary for the present study briefly here.

The completeness of the eigenstates of positions and spins for $N$-electron systems is expressed as
\begin{gather}
	\frac{1}{N !}
    \int
   	d \xi_1
	\cdots
    d \xi_N
       | \xi_1, \dots, \xi_N \rangle
        \langle \xi_1, \dots, \xi_N |
	=
    	\hat{1}
	,
	\label{identity_for_N_els}        
\end{gather}
where $\xi$ represents the position and spin variables collectively.
For an $N$-electron state $| \Psi^{(N)} \rangle$ normalized to unity ($\langle \Psi^{(N)} | \Psi^{(N)} \rangle = 1$),
the completeness in eq. (\ref{identity_for_N_els}) imposes the normalization condition on the many-electron WF as
\begin{gather}
    \int
    d \xi_1
	\cdots
	d \xi_N
    	|\Psi^{(N)} (\xi_1, \dots , \xi_N)|^2    	
	=
    	N !
	.
	\label{integ_sq_many_el_WF}
\end{gather}
The electron density corresponding to a many-electron WF is calculated as
\begin{gather}
	n (\xi)
    =
    	\frac{1}{(N - 1) !}
	    \int
   		d \xi_1
		\cdots
		d \xi_{N - 1}
	    	|\Psi^{(N)} (\xi_1, \dots , \xi_{N - 1}, \xi)|^2    	
	,
	\label{el_dens_from_WF}
\end{gather}
which gives, as expected, the total electron number:
$\int d \xi n (\xi) = N$.

By referring to the action of the creation operator of the electron field\cite{stefanucci2013nonequilibrium},
we can express the matrix element of the operator as
\begin{gather}
	\langle \Psi^{(N + 1)} |
    \hat{\psi}^\dagger (\xi)
    | \Psi^{(N)} \rangle
    =
		\frac{1}{N !}
        \int
        d \xi_1 \cdots
        d \xi_N
	\cdot
	\nonumber \\
	\cdot
			\Psi^{(N + 1)} (\xi_1, \dots, \xi_N, \xi)^*
            \Psi^{(N)} (\xi_1, \dots, \xi_N)
	.
	\label{matel_creation_opr_as_integ_of_WFs}
\end{gather}
The expression for the annihilation operator can be obtained by taking the complex conjugate of eq. (\ref{matel_creation_opr_as_integ_of_WFs}).

\section{proof of eq. (\ref{phi123_phi231_phi312_simple})}
\label{appendix:proof_of_phi123}

From eqs. (\ref{X_in_x1_x2_x3}), (\ref{x_in_x1_x2}), and (\ref{y_in_x1_x2_x3}),
we have
\begin{gather}
	x_1 - x_2
    =
    	\sqrt{2} r \sin \phi
	\nonumber        
	\\
	x_2 - x_3
    =
    	\sqrt{2} r \sin \left( \phi + \frac{2 \pi}{3} \right)
	\nonumber        
	\\
	x_3 - x_1
    =
    	\sqrt{2} r \sin \left( \phi + \frac{4 \pi}{3} \right)
	.
	\label{x_diffs_in_r_phi}        
\end{gather}
The angle variable $\phi (x_1, x_2, x_3) \equiv \phi(123)$ falls within the following ranges according to $x_1, x_2$, and $x_3$ [see eq. (3.8) in Ref. \cite{bib:3935}]:
\begin{gather}
	\begin{cases}
    	0 < \phi (123) < \pi/3 & (x_1 > x_2 > x_3) \\
    	\pi/3 < \phi (123) < 2 \pi/3 & (x_1 > x_3 > x_2) \\
    	2 \pi/3 < \phi (123) < \pi & (x_3 > x_1 > x_2) \\
    	\pi < \phi (123) < 4 \pi/3 & (x_3 > x_2 > x_1) \\
    	4 \pi/3 < \phi (123) < 5 \pi/3 & (x_2 > x_3 > x_1) \\
    	5 \pi/3 < \phi (123) < 2 \pi & (x_2 > x_1 > x_3) \\
	\end{cases}
    .
    \label{phi_range_for_x1_x2_x3}
\end{gather}
The ranges of $\phi(231)$ and $\phi(312)$ are restricted similarly.
From eq. (\ref{x_diffs_in_r_phi}),
we obtain the relations between $\phi (123)$, $\phi(231)$, and $\phi (312)$ as
\begin{gather}
	\tan \phi (231)
	=
		\frac{\sqrt{3} (x_2 - x_3)}{x_2 + x_3 - 2 x_1}
	=
    	\tan
        \left[ \phi (123) + \frac{2 \pi}{3} \right]
	.
\end{gather}
and
\begin{gather}
	\tan \phi (312)
	=
        \frac{\sqrt{3} (x_3 - x_1)}{x_3 + x_1 - 2 x_2}
	=
    	\tan
        \left[ \phi (123) + \frac{4 \pi}{3} \right]
	.
\end{gather}
With these relations and eq. (\ref{phi_range_for_x1_x2_x3}),
the three angle variables are found to satisfy eq. (\ref{phi123_phi231_phi312_simple}).

\section{non-interacting case for three electrons}
\label{appendix:non_int_3e}

For a non-interacting case $(\Lambda = 0)$,
the spatial dynamics of the three-electron systems is described as that of three independent harmonic oscillators.
The energy eigenvalue of the whole system is thus given by
$
E^{(3) \mathrm{non-int}}_{n_1 n_2 n_3}
=
\omega_0
\left( n_1 + n_2 + n_3 + 3/2 \right)
,
$
where $n_1, n_2$, and $n_3$ are the quantum numbers for the independent oscillators.
Since the three electrons are equivalent,
it is clear that the electronic properties of the system are determined not by the order of the quantum numbers, but by the combination of them.

\subsection{$n_1 = n_2 = n_3$ case}

Any electronic configuration in which the three quantum numbers are equal to each other is impossible due to the Fermi statistics.

\subsection{$n_1 \ne n_2 = n_3$ case}

For $n_1 \equiv n \ne n_2 = n_3 \equiv n'$,
the spatial WF
$
\psi_{n n' n'} (x_1, x_2, x_3)
\equiv
\psi_{n} (x_1) \psi_{n'} (x_2) \psi_{n'} (x_3)
$
vanishes when it is anti-symmetrized:
$\mathcal{A} \psi_{n n' n'} (x_1, x_2, x_3) = 0$,
indicating that the construction of an $S = 3/2$ state from this spatial WF is impossible [see eq. (\ref{antisymmetrized_f_spin_32})].
On the other hand, 
by using eq. (\ref{antisymmetrized_f_spin_half}) with $j = 1$,
we can construct the anti-symmetrized three-electron WF for an $S = 1/2$ state as
\begin{widetext}
\begin{gather}
	\psi_{n n' n'}^{1/2, S_z, 1}
    (x_1, s_1, x_2, s_2, x_3, s_3)
	\nonumber \\
	=
    	\sqrt{\frac{3}{2}}
    	\chi_1^{1/2, S_z} (123)
        \left[
            \psi_{n n' n'} (123)
            -
            \psi_{n n' n'} (231)
        \right]
        +
        \frac{1}{\sqrt{2}}
    	\chi_2^{1/2, S_z} (123)
        [
        	-
            \psi_{n n' n'} (123)
            -
            \psi_{n n' n'} (231)
            +
        	2
            \psi_{n n' n'} (312)
        ]
	,
\end{gather}
where we introduced a normalization constant.
Although the anti-symmetrized WF $\psi_{n n' n'}^{1/2, S_z, 2}$
from $\psi_{n n' n'}$ for $j = 2$ can be constructed as well,
one can confirm that it is the same WF as $\psi_{n n' n'}^{1/2, S_z, 1}$.
The simultaneous eigenstates of the energy, $S = 1/2$,
and $S_z$ are thus covered by $\psi_{n n' n'}^{1/2, S_z, 1}$:
\begin{gather}
	\psi_{n n' n'}^{1/2, S_z}
    (x_1, s_1, x_2, s_2, x_3, s_3)
    \equiv
    	\psi_{n n' n'}^{1/2, S_z, 1} 
	    (x_1, s_1, x_2, s_2, x_3, s_3)
	,        
	\label{eigen_func_non_int_n1_n2_n2}
\end{gather}
implying that the energy eigenstates for the combination $\{ n, n', n' \}$
of quantum numbers are doubly degenerate.

\subsection{$n_1 \ne n_2, n_2 \ne n_3, n_3 \ne n_1$ case}

By using eq. (\ref{antisymmetrized_f_spin_32}) for the spatial function
$
\psi_{n_1 n_2 n_3} (x_1, x_2, x_3) 
\equiv
\psi_{n_1} (x_1) \psi_{n_2} (x_2) \psi_{n_3} (x_3)
,
$
we can construct the anti-symmetrized three-electron WF for an $S = 3/2$ state as
\begin{gather}
	\psi_{n_1 n_2 n_3}^{3/2, S_z} (123)
	\nonumber \\
	=
		\chi^{3/2, S_z} (123)
        [
        	\psi_{n_1 n_2 n_3} (123)
            -
        	\psi_{n_1 n_2 n_3} (213)
            -
        	\psi_{n_1 n_2 n_3} (132)
            -
        	\psi_{n_1 n_2 n_3} (321)
            +
        	\psi_{n_1 n_2 n_3} (231)
            +
        	\psi_{n_1 n_2 n_3} (312)
        ]
	.        
\end{gather}

By using eq. (\ref{antisymmetrized_f_spin_half}) with $j = 1, 2$,
we can construct the anti-symmetrized three-electron WF for $S = 1/2$ states:
\begin{gather}
	\psi_{n_1 n_2 n_3}^{1/2, S_z, 1} (123)
	\nonumber \\
	=
	    \frac{1}{2}
		\chi^{1/2, S_z}_1 (123)
        \Bigg[
        	2
        	\psi_{n_1 n_2 n_3} (123)
            -
            2
        	\psi_{n_1 n_2 n_3} (213)
            +
            \psi_{n_1 n_2 n_3} (132)
            +
            \psi_{n_1 n_2 n_3} (321)
            -
            \psi_{n_1 n_2 n_3} (231)
            -
            \psi_{n_1 n_2 n_3} (312)
        \Bigg]
	\nonumber \\
    	+
        \frac{\sqrt{3}}{2}
		\chi^{1/2, S_z}_2 (123)
        \left[
        	-
            \psi_{n_1 n_2 n_3} (132)
            +
            \psi_{n_1 n_2 n_3} (321)
        	-
            \psi_{n_1 n_2 n_3} (231)
            +
            \psi_{n_1 n_2 n_3} (312)
        \right]
\end{gather}
and
\begin{gather}
	\psi_{n_1 n_2 n_3}^{1/2, S_z, 2} (123)
	\nonumber \\
	=
        \frac{\sqrt{3}}{2}
		\chi^{1/2, S_z}_1 (123)
        \left[
            -
        	\psi_{n_1 n_2 n_3} (132)
            +
            \psi_{n_1 n_2 n_3} (321)
            +
            \psi_{n_1 n_2 n_3} (231)
            -
            \psi_{n_1 n_2 n_3} (312)
        \right]
	\nonumber \\
    	+
        \frac{1}{2}
		\chi^{1/2, S_z}_2 (123)
        \left[
        	2
        	\psi_{n_1 n_2 n_3} (123)
            +
            2
        	\psi_{n_1 n_2 n_3} (213)
        	-
        	\psi_{n_1 n_2 n_3} (132)
        	-
        	\psi_{n_1 n_2 n_3} (321)
        	-
        	\psi_{n_1 n_2 n_3} (231)
        	-
        	\psi_{n_1 n_2 n_3} (312)
        \right]
	,
\end{gather}
\end{widetext}
which are easily confirmed to be orthogonal to each other.
The energy eigenstates for the combination $\{ n_1, n_2, n_3 \}$
of quantum numbers are eight-fold degenerate.

Some of the lowest energy eigenvalues and the corresponding three-electron WFs, the total spin,
and the degeneracy are shown in Table \ref{table_non_int_energy}.
The ground states are doubly degenerate $S = 1/2$ states,
corresponding to $\{ n_1, n_2, n_3 \} = \{1, 0, 0 \}$.

\begin{table}
\centering
\caption{Some of the lowest eigenvalues $E^{(3) \mathrm{non-int}}_{n_1 n_2 n_3}$ of the non-interacting three-electron system. The three-electron wave functions, the total spin $S$, and the degeneracy are also shown.}
\label{table_non_int_energy}
\begin{tabular}{ccc} 
\hline
$E^{(3) \mathrm{non-int}}_{n_1 n_2 n_3}/\omega_0$ & $\psi_{n_1 n_2 n_3}^{S, S_z}$ & degeneracy  \\ 
\hline
$5/2$                                             & $\psi_{100}^{1/2, S_z}$       & 2           \\
$7/2$                                             & $\psi_{110}^{1/2, S_z}$       & 2           \\
\multicolumn{1}{l}{}                              & $\psi_{200}^{1/2, S_z}$       & 2           \\
$9/2$                                             & $\psi_{210}^{3/2, S_z}$       & 4           \\
\multicolumn{1}{l}{}                              & $\psi_{210}^{1/2, S_z, 1}$    & 2           \\
\multicolumn{1}{l}{}                              & $\psi_{210}^{1/2, S_z, 2}$    & 2           \\
\multicolumn{1}{l}{}                              & $\psi_{300}^{1/2, S_z}$       & 2           \\
\hline
\end{tabular}
\end{table}

\section{calculation of $M_j^{S_z} (s_3)$}
\label{appendix:calc_M_spin}

The spin WF of a singlet state for two spins is given by
$
\chi^{0,0} (s_1, s_2)
=
[ \alpha (s_1) \beta (s_2) - \beta (s_1) \alpha (s_2) ]/\sqrt{2}
$
.
By substituting this expression and
eqs. (\ref{spin_wf_doublet_1_up}) and (\ref{spin_wf_doublet_1_dn})
into eq. (\ref{def_matel_spin_s3}) for $j = 1$
and using the orthonormality of the spin WFs,
we can calculate the integrals as
$
	M_1^{1/2} (s_3)
    =
		M_1^{-1/2} (s_3)
    =
    	0
	.
$
We can also calculate those for $j = 2$ from
eqs. (\ref{spin_wf_doublet_2_up}) and (\ref{spin_wf_doublet_2_dn}) as
$
M_2^{1/2} (s_3)
    =
		\alpha (s_3)^*
$
and
$
M_2^{-1/2} (s_3)
    =
    	-\beta (s_3)^*
	.
$

\section{calculation of $M_{k m n}^+ (x_3)$}
\label{appendix:calc_M_spatial}

\begin{figure}
\begin{center}
\includegraphics[width=5.5cm]{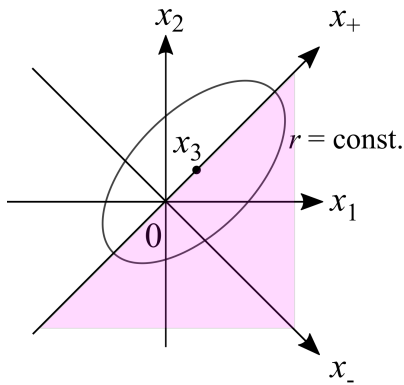}
\end{center}
\caption{
For a given $r$, allowed $x_+$ and $x_-$ form an ellipse centered at $(x_3, 0)$ on the $x_+$-$x_-$ plane.
For the integral over the region of $x_1 > x_2$, shaded in the figure,
the range of integration variable $r$ for a fixed $x_+$ is $\sqrt{2/3} |x_+ - x_3| < r < \infty$.
}
\label{Fig_coord_for_M}
\end{figure}

By substituting the expressions for the many-electron WFs in eqs. (\ref{harm_delta_two_el_wave_func}) and (\ref{spatial_WF_3e}) into the definition eq. (\ref{def_matel_spatial_x3}) of the integral,
we obtain
\begin{widetext}
\begin{gather}
	M_{k n m}^+ (x_3)
    =
    	\sqrt{\frac{m_e \omega_0}{\pi}}
        (4 \lambda_2)^{1/4}
		\int
    	d x_1
    	d x_2
	    	\psi^{\mathrm{HO}}_k
    	    \left(
                \frac{2 x_+ + x_3}{\sqrt{3} \ell};
                \frac{\ell}{\sqrt{3}}
			\right)
            R_{n m} (r)
	        \Phi_m^+ (\phi )
            \exp
            \left[
            	-m_e \omega_0
                \left( x_+^2 + \frac{\lambda_2 x_-^2}{2}  \right)
			\right]
	.
    \label{integ_Psi3_kmn_Psi2_00}
\end{gather}
\end{widetext}
To proceed the calculation, we consider the transformation of the integral variables from $(x_1, x_2)$ to $(x_+, r)$.
It is confirmed that from eqs. (\ref{x_in_x1_x2})-(\ref{def_r_in_x_y})
that $r, x_+$, and $x_-$ are related via the relation
\begin{gather}
	r^2
	=
    	x_-^2
        +
        \frac{2}{3}
        (x_+ - x_3)^2
	,
	\label{r2_in_X2_x2_param_x3}        
\end{gather}
which represents an ellipse centered at $(x_3, 0)$ on the $x_+$-$x_-$ plane for a given $r$.
$x_1$ and $x_2$ are expressed in $x_+$ and $x_-$ as
$x_1 = x_+ + x_-/\sqrt{2}$
and
$x_2 = x_+ - x_-/\sqrt{2}$, respectively.
Since the integrand in eq. (\ref{integ_Psi3_kmn_Psi2_00}) is invariant under the exchange of $x_1$ and $x_2$,
it is necessary to consider only the case with $x_1 > x_2$,
in which we can express $x_1$ and $x_2$ as functions of $x_+$ and $r$ as
\begin{gather}
	x_{1, 2}
    =
    	x_+
        \pm
        \sqrt{\frac{r^2}{2} - \frac{(x_+ - x_3)^2}{3}}
	.
\end{gather}
The positive (negative) sign on the right-hand side is for $x_1 \, (x_2)$.
The Jacobian for the transformation of integral variables is then calculated as
\begin{gather}
	\frac{\partial (x_1, x_2)}{\partial (x_+, r)}
	=
    	-
        \left[
        	\frac{1}{2} - \frac{(x_+ - x_3)^2}{3 r^2}
		\right]^{-1/2}
	.
\end{gather}
For a given $x_+$, it is clear from eq. (\ref{r2_in_X2_x2_param_x3})
that the range of $r$ is $\sqrt{2/3} |x_+ - x_3| < r < \infty$ to represent $0 < x_- < \infty$.
The integral measure in eq. (\ref{integ_Psi3_kmn_Psi2_00}) is rewritten via the transformation as
\begin{gather}
	\int
    d x_1
    d x_2
    =
    	2
    	\int_{x_1 > x_2}
        d x_1
        d x_2
	\nonumber \\
	=
		2
        \int_{-\infty}^\infty
        d x_+
        \int_{\sqrt{2/3} |x_+ - x_3|}^\infty
        d r
		\left|
        	\frac{\partial (x_1, x_2)}{\partial (x_+, r)}
		\right|
	,
\end{gather}
for which the coordinate systems are shown in Fig. \ref{Fig_coord_for_M}. 
By using the transformation and introducing the dimensionless coordinates
$\overline{x}_+ \equiv \sqrt{3} (x_+ - x_3)/\ell$
and
$\overline{x}_3 \equiv \sqrt{3} x_3/\ell$,
we can rewrite eq. (\ref{integ_Psi3_kmn_Psi2_00}) as
\begin{widetext}
\begin{gather}
	M_{k n m}^+ (x_3)
    =
    	\left( \frac{\lambda_2}{3} \right)^{1/4}
        \sqrt{\frac{32 n!}{\pi (n + m) !}}
		\int_{-\infty}^\infty
			d \overline{x}_+
            \psi^{\mathrm{HO}}_k
    	    \left(
                \frac{2}{3} \overline{x}_+ + \overline{x}_3;
                \frac{\ell}{\sqrt{3}}
			\right)
            \exp
            \left[
            	-\frac{ (\overline{x}_+ + \overline{x}_3)^2 }{6}
                +
	            \frac{\lambda_2}{18}
                \overline{x}_+^2
			\right]
	\cdot
    \nonumber \\
    \cdot
		\int_{\sqrt{\lambda_3} | \overline{x}_+ |/3}^\infty
        	d \overline{r}
            \frac{
            	\overline{r}^m
                L_n^m (\overline{r}^2)
                \Phi_m^+ (\phi)}
                {\sqrt{1 - \lambda_3 \overline{x}_+^2/ (9 \overline{r}^2) }}
	    	\exp
            \left[
            	-\frac{\overline{r}^2}{2}
                \left( 1 + \frac{\lambda_2}{\lambda_3} \right)
            \right]
	=
        \left( \frac{m_e \omega_0 \lambda_2}{\pi^5} \right)^{1/4}
		\sqrt{\frac{32 n !}{2^k k! (n + m)!}}
        \widetilde{M}_{k n m} (\overline{x}_3)
	,
	\label{matel_M_in_tilde_M}
\end{gather}
where we used the expression of $R_{n m} (r)$ in eq. (\ref{solution_R_n_m})
and used the relation $\cos \phi = y/r = \overline{x}_+ \sqrt{\lambda_3}/(3 \overline{r})$.
We defined the dimensionless integral
\begin{gather}
	\widetilde{M}_{k n m} (\overline{x})
	=
		\int_{-\infty}^\infty
			d \zeta
            \exp
         	\left[
            	-\frac{1}{4}
				\left(
                	\frac{2}{3} \zeta + \overline{x}
				\right)^2
            		-\frac{ (\zeta + \overline{x})^2 }{6}
			\right]
            H_k
            \left(
            	\frac{1}{\sqrt{2}}
                \left\{
                	\frac{2}{3} \zeta + \overline{x}
				\right\}  
			\right)
		I_{n m}
    	\left(
    		\frac{\sqrt{\lambda_3}}{3}
        	\zeta 
		\right)
	\label{def_matel_tilde_M}        
\end{gather}
with
\begin{gather}
	I_{n m} (\overline{r}_0)
    \equiv
		\int_{| \overline{r}_0 |}^\infty
		d \overline{r}
        	\frac{\overline{r}^m}
            {\sqrt{1 - (\overline{r}_0/ \overline{r})^2}}
        	L_n^m (\overline{r}^2)
	        T_m \left( \frac{\overline{r}_0}{\overline{r}} \right)
            \exp
            \left[
            	-\frac{\overline{r}^2}{2}
                -\frac{\lambda_2}{2 \lambda_3}
                (\overline{r}^2 - \overline{r}_0^2)
            \right]
	\nonumber \\
	=
		-
		( \mathrm{sgn} \, \overline{r}_0 )
		\overline{r}_0^{m + 1}
		\int_1^\infty
		d \zeta
	        \sqrt{\zeta^2 - 1}
            \frac{d}{d \zeta}
            \left[
        	\zeta^m
        	L_n^m (\overline{r}_0^2 \zeta^2)
	        T_m \left( \frac{1}{\zeta} \right)
            \exp
            \left\{
            	-\frac{\overline{r}_0^2}{2}
                \zeta^2
                -\frac{\lambda_2}{2 \lambda_3}
                \overline{r}_0^2
                (\zeta^2 - 1)
            \right\}
			\right]
	,            
	\label{def_integ_I_n_m_r0}          
\end{gather}
where we used partial integration.
$T_m$ is the Chebyshev polynomial of the first kind,
defined so that $T_m (\cos \phi) = \cos (m \phi)$.

\end{widetext}

\section{Hole excitation of three-electron system}
\label{appendix:hole_excitation_3e}

Since we have the exact WFs for the two- and three-electron systems in hand,
we can derive the analytic expressions for the interacting GF of the three-electron system for hole excitation.
Here we discuss briefly their pole positions and the relation with the spectral intensities.

\begin{figure}
\begin{center}
\includegraphics[width=7cm]{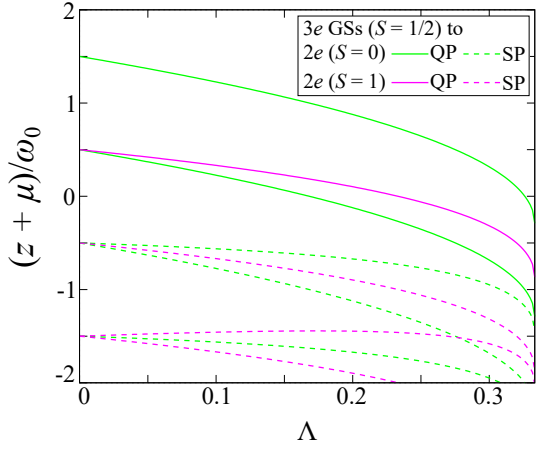}
\end{center}
\caption{
Pole positions of the interacting GF of the three-electron system for hole excitation as
functions of the interaction strength $\Lambda$.
The quasiparticle peaks (QPs) and the satellite peaks (SPs) are distinguished in the plot
for the transitions from the three-electron ground states (GSs) to the two-electron states.
}
\label{Fig_poles_3e_hole}
\end{figure}

By using the WFs in eqs. (\ref{WF_2e_even_nr}) and (\ref{WF_2e_odd_nr}) for the two-electron system,
we can calculate the partial GFs
$G^{(\mathrm{h}) \to \mathrm{singlet}}$ and
$G^{(\mathrm{h}) \to \mathrm{triplet}}$
for the transitions from the three-electron ground states to the spin singlet and triplet two-electron states, respectively.
Specifically, the spectral representations of
$G^{(\mathrm{h}) \to \mathrm{singlet}}$ and
$G^{(\mathrm{h}) \to \mathrm{triplet}}$ involve the infinite summation over even and odd $n_{\mathrm{r}}$, respectively,
while each of them involves that over all non-negative integers for $n_{\mathrm{c}}$.
We can thus use the formula\cite{bib:3963} for the summation over $n_{\mathrm{c}}$ to express 
$G^{(\mathrm{h}) \to \mathrm{singlet}}$ and 
$G^{(\mathrm{h}) \to \mathrm{triplet}}$ using the parabolic cylinder functions similarly to the GF of the two-electron system.
One can easily find their expressions contain the Gamma functions
whose poles are given by
\begin{gather}
	\frac{z + \mu}{\omega_0}
	-
    2
    \lambda_3
    +
    \lambda_2
    \left( n_{\mathrm{r}} + \frac{1}{2} \right)
	=
		0, -1, -2, \dots,
\end{gather}
which are nothing but the poles of $G^{(\mathrm{h})}$.
The pole positions as functions of the interaction strength is plotted in Fig. \ref{Fig_poles_3e_hole}.
By comparing the two-electron WFs and the one-electron orbitals,
it is found that the highest pole on $z$ axis corresponds to the HOMO for the unpaired electron in the three-electron system
and the second and third highest ones correspond to the occupied lowest-energy orbitals.
These three poles should thus be identified as the quasiparticle peaks in the spectral function.
The other poles do not have corresponding one-electron orbitals
and they are identified as the satellite peaks.
The distance between the pole positions of the second and third highest quasiparticle peaks are nothing but the exchange splitting,
which is easily calculated as
$\Delta_{\mathrm{x}} = \omega_0 (1 - \lambda_2)$.

\bibliographystyle{apsrev4-1}
\bibliography{paper}

\begin{thebibliography}{40}%
\makeatletter
\providecommand \@ifxundefined [1]{%
 \@ifx{#1\undefined}
}%
\providecommand \@ifnum [1]{%
 \ifnum #1\expandafter \@firstoftwo
 \else \expandafter \@secondoftwo
 \fi
}%
\providecommand \@ifx [1]{%
 \ifx #1\expandafter \@firstoftwo
 \else \expandafter \@secondoftwo
 \fi
}%
\providecommand \natexlab [1]{#1}%
\providecommand \enquote  [1]{``#1''}%
\providecommand \bibnamefont  [1]{#1}%
\providecommand \bibfnamefont [1]{#1}%
\providecommand \citenamefont [1]{#1}%
\providecommand \href@noop [0]{\@secondoftwo}%
\providecommand \href [0]{\begingroup \@sanitize@url \@href}%
\providecommand \@href[1]{\@@startlink{#1}\@@href}%
\providecommand \@@href[1]{\endgroup#1\@@endlink}%
\providecommand \@sanitize@url [0]{\catcode `\\12\catcode `\$12\catcode
  `\&12\catcode `\#12\catcode `\^12\catcode `\_12\catcode `\%12\relax}%
\providecommand \@@startlink[1]{}%
\providecommand \@@endlink[0]{}%
\providecommand \url  [0]{\begingroup\@sanitize@url \@url }%
\providecommand \@url [1]{\endgroup\@href {#1}{\urlprefix }}%
\providecommand \urlprefix  [0]{URL }%
\providecommand \Eprint [0]{\href }%
\providecommand \doibase [0]{http://dx.doi.org/}%
\providecommand \selectlanguage [0]{\@gobble}%
\providecommand \bibinfo  [0]{\@secondoftwo}%
\providecommand \bibfield  [0]{\@secondoftwo}%
\providecommand \translation [1]{[#1]}%
\providecommand \BibitemOpen [0]{}%
\providecommand \bibitemStop [0]{}%
\providecommand \bibitemNoStop [0]{.\EOS\space}%
\providecommand \EOS [0]{\spacefactor3000\relax}%
\providecommand \BibitemShut  [1]{\csname bibitem#1\endcsname}%
\let\auto@bib@innerbib\@empty
%</preamble>
\bibitem [{\citenamefont {Kestner}\ and\ \citenamefont
  {Sinano\ifmmode~\bar{g}\else \={g}\fi{}lu}(1962)}]{bib:2817}%
  \BibitemOpen
  \bibfield  {author} {\bibinfo {author} {\bibfnamefont {N.~R.}\ \bibnamefont
  {Kestner}}\ and\ \bibinfo {author} {\bibfnamefont {O.}~\bibnamefont
  {Sinano\ifmmode~\bar{g}\else \={g}\fi{}lu}},\ }\href {\doibase
  10.1103/PhysRev.128.2687} {\bibfield  {journal} {\bibinfo  {journal} {Phys.
  Rev.}\ }\textbf {\bibinfo {volume} {128}},\ \bibinfo {pages} {2687} (\bibinfo
  {year} {1962})}\BibitemShut {NoStop}%
\bibitem [{\citenamefont {Taut}(1993)}]{bib:906}%
  \BibitemOpen
  \bibfield  {author} {\bibinfo {author} {\bibfnamefont {M.}~\bibnamefont
  {Taut}},\ }\href {\doibase 10.1103/PhysRevA.48.3561} {\bibfield  {journal}
  {\bibinfo  {journal} {Phys. Rev. A}\ }\textbf {\bibinfo {volume} {48}},\
  \bibinfo {pages} {3561} (\bibinfo {year} {1993})}\BibitemShut {NoStop}%
\bibitem [{\citenamefont {Taut}(1994)}]{bib:3926}%
  \BibitemOpen
  \bibfield  {author} {\bibinfo {author} {\bibfnamefont {M.}~\bibnamefont
  {Taut}},\ }\href {http://stacks.iop.org/0305-4470/27/i=3/a=040} {\bibfield
  {journal} {\bibinfo  {journal} {Journal of Physics A: Mathematical and
  General}\ }\textbf {\bibinfo {volume} {27}},\ \bibinfo {pages} {1045}
  (\bibinfo {year} {1994})}\BibitemShut {NoStop}%
\bibitem [{\citenamefont {Merkt}\ \emph {et~al.}(1991)\citenamefont {Merkt},
  \citenamefont {Huser},\ and\ \citenamefont {Wagner}}]{bib:4475}%
  \BibitemOpen
  \bibfield  {author} {\bibinfo {author} {\bibfnamefont {U.}~\bibnamefont
  {Merkt}}, \bibinfo {author} {\bibfnamefont {J.}~\bibnamefont {Huser}}, \ and\
  \bibinfo {author} {\bibfnamefont {M.}~\bibnamefont {Wagner}},\ }\href
  {\doibase 10.1103/PhysRevB.43.7320} {\bibfield  {journal} {\bibinfo
  {journal} {Phys. Rev. B}\ }\textbf {\bibinfo {volume} {43}},\ \bibinfo
  {pages} {7320} (\bibinfo {year} {1991})}\BibitemShut {NoStop}%
\bibitem [{\citenamefont {Guan}\ \emph {et~al.}(2013)\citenamefont {Guan},
  \citenamefont {Batchelor},\ and\ \citenamefont {Lee}}]{bib:4492}%
  \BibitemOpen
  \bibfield  {author} {\bibinfo {author} {\bibfnamefont {X.-W.}\ \bibnamefont
  {Guan}}, \bibinfo {author} {\bibfnamefont {M.~T.}\ \bibnamefont {Batchelor}},
  \ and\ \bibinfo {author} {\bibfnamefont {C.}~\bibnamefont {Lee}},\ }\href
  {\doibase 10.1103/RevModPhys.85.1633} {\bibfield  {journal} {\bibinfo
  {journal} {Rev. Mod. Phys.}\ }\textbf {\bibinfo {volume} {85}},\ \bibinfo
  {pages} {1633} (\bibinfo {year} {2013})}\BibitemShut {NoStop}%
\bibitem [{\citenamefont {Hohenberg}\ and\ \citenamefont
  {Kohn}(1964)}]{bib:76}%
  \BibitemOpen
  \bibfield  {author} {\bibinfo {author} {\bibfnamefont {P.}~\bibnamefont
  {Hohenberg}}\ and\ \bibinfo {author} {\bibfnamefont {W.}~\bibnamefont
  {Kohn}},\ }\href {\doibase 10.1103/PhysRev.136.B864} {\bibfield  {journal}
  {\bibinfo  {journal} {Phys. Rev.}\ }\textbf {\bibinfo {volume} {136}},\
  \bibinfo {pages} {B864} (\bibinfo {year} {1964})}\BibitemShut {NoStop}%
\bibitem [{\citenamefont {Kohn}\ and\ \citenamefont {Sham}(1965)}]{bib:77}%
  \BibitemOpen
  \bibfield  {author} {\bibinfo {author} {\bibfnamefont {W.}~\bibnamefont
  {Kohn}}\ and\ \bibinfo {author} {\bibfnamefont {L.~J.}\ \bibnamefont
  {Sham}},\ }\href {\doibase 10.1103/PhysRev.140.A1133} {\bibfield  {journal}
  {\bibinfo  {journal} {Phys. Rev.}\ }\textbf {\bibinfo {volume} {140}},\
  \bibinfo {pages} {A1133} (\bibinfo {year} {1965})}\BibitemShut {NoStop}%
\bibitem [{\citenamefont {Hedin}(1965)}]{bib:GW1}%
  \BibitemOpen
  \bibfield  {author} {\bibinfo {author} {\bibfnamefont {L.}~\bibnamefont
  {Hedin}},\ }\href {\doibase 10.1103/PhysRev.139.A796} {\bibfield  {journal}
  {\bibinfo  {journal} {Phys. Rev.}\ }\textbf {\bibinfo {volume} {139}},\
  \bibinfo {pages} {A796} (\bibinfo {year} {1965})}\BibitemShut {NoStop}%
\bibitem [{\citenamefont {Hybertsen}\ and\ \citenamefont
  {Louie}(1985)}]{bib:GW2}%
  \BibitemOpen
  \bibfield  {author} {\bibinfo {author} {\bibfnamefont {M.~S.}\ \bibnamefont
  {Hybertsen}}\ and\ \bibinfo {author} {\bibfnamefont {S.~G.}\ \bibnamefont
  {Louie}},\ }\href {\doibase 10.1103/PhysRevLett.55.1418} {\bibfield
  {journal} {\bibinfo  {journal} {Phys. Rev. Lett.}\ }\textbf {\bibinfo
  {volume} {55}},\ \bibinfo {pages} {1418} (\bibinfo {year}
  {1985})}\BibitemShut {NoStop}%
\bibitem [{\citenamefont {Hybertsen}\ and\ \citenamefont
  {Louie}(1986)}]{bib:GW3}%
  \BibitemOpen
  \bibfield  {author} {\bibinfo {author} {\bibfnamefont {M.~S.}\ \bibnamefont
  {Hybertsen}}\ and\ \bibinfo {author} {\bibfnamefont {S.~G.}\ \bibnamefont
  {Louie}},\ }\href {\doibase 10.1103/PhysRevB.34.5390} {\bibfield  {journal}
  {\bibinfo  {journal} {Phys. Rev. B}\ }\textbf {\bibinfo {volume} {34}},\
  \bibinfo {pages} {5390} (\bibinfo {year} {1986})}\BibitemShut {NoStop}%
\bibitem [{\citenamefont {Damascelli}(2004)}]{bib:4070}%
  \BibitemOpen
  \bibfield  {author} {\bibinfo {author} {\bibfnamefont {A.}~\bibnamefont
  {Damascelli}},\ }\href {http://stacks.iop.org/1402-4896/2004/i=T109/a=005}
  {\bibfield  {journal} {\bibinfo  {journal} {Physica Scripta}\ }\textbf
  {\bibinfo {volume} {2004}},\ \bibinfo {pages} {61} (\bibinfo {year}
  {2004})}\BibitemShut {NoStop}%
\bibitem [{\citenamefont {Moser}(2017)}]{bib:4165}%
  \BibitemOpen
  \bibfield  {author} {\bibinfo {author} {\bibfnamefont {S.}~\bibnamefont
  {Moser}},\ }\href {\doibase https://doi.org/10.1016/j.elspec.2016.11.007}
  {\bibfield  {journal} {\bibinfo  {journal} {Journal of Electron Spectroscopy
  and Related Phenomena}\ }\textbf {\bibinfo {volume} {214}},\ \bibinfo {pages}
  {29 } (\bibinfo {year} {2017})}\BibitemShut {NoStop}%
\bibitem [{\citenamefont {Kosugi}\ \emph {et~al.}(2017)\citenamefont {Kosugi},
  \citenamefont {Nishi}, \citenamefont {Kato},\ and\ \citenamefont
  {Matsushita}}]{bib:pw_unfolding}%
  \BibitemOpen
  \bibfield  {author} {\bibinfo {author} {\bibfnamefont {T.}~\bibnamefont
  {Kosugi}}, \bibinfo {author} {\bibfnamefont {H.}~\bibnamefont {Nishi}},
  \bibinfo {author} {\bibfnamefont {Y.}~\bibnamefont {Kato}}, \ and\ \bibinfo
  {author} {\bibfnamefont {Y.-i.}\ \bibnamefont {Matsushita}},\ }\href
  {\doibase 10.7566/JPSJ.86.124717} {\bibfield  {journal} {\bibinfo  {journal}
  {Journal of the Physical Society of Japan}\ }\textbf {\bibinfo {volume}
  {86}},\ \bibinfo {pages} {124717} (\bibinfo {year} {2017})},\ \Eprint
  {http://arxiv.org/abs/https://doi.org/10.7566/JPSJ.86.124717}
  {https://doi.org/10.7566/JPSJ.86.124717} \BibitemShut {NoStop}%
\bibitem [{\citenamefont {Furukawa}\ \emph {et~al.}(2018)\citenamefont
  {Furukawa}, \citenamefont {Kosugi}, \citenamefont {Nishi},\ and\
  \citenamefont {Matsushita}}]{bib:4473}%
  \BibitemOpen
  \bibfield  {author} {\bibinfo {author} {\bibfnamefont {Y.}~\bibnamefont
  {Furukawa}}, \bibinfo {author} {\bibfnamefont {T.}~\bibnamefont {Kosugi}},
  \bibinfo {author} {\bibfnamefont {H.}~\bibnamefont {Nishi}}, \ and\ \bibinfo
  {author} {\bibfnamefont {Y.-i.}\ \bibnamefont {Matsushita}},\ }\href
  {\doibase 10.1063/1.5029537} {\bibfield  {journal} {\bibinfo  {journal} {The
  Journal of Chemical Physics}\ }\textbf {\bibinfo {volume} {148}},\ \bibinfo
  {pages} {204109} (\bibinfo {year} {2018})},\ \Eprint
  {http://arxiv.org/abs/https://doi.org/10.1063/1.5029537}
  {https://doi.org/10.1063/1.5029537} \BibitemShut {NoStop}%
\bibitem [{\citenamefont {Kosugi}\ \emph {et~al.}(2018)\citenamefont {Kosugi},
  \citenamefont {Nishi}, \citenamefont {Furukawa},\ and\ \citenamefont
  {Matsushita}}]{bib:4483}%
  \BibitemOpen
  \bibfield  {author} {\bibinfo {author} {\bibfnamefont {T.}~\bibnamefont
  {Kosugi}}, \bibinfo {author} {\bibfnamefont {H.}~\bibnamefont {Nishi}},
  \bibinfo {author} {\bibfnamefont {Y.}~\bibnamefont {Furukawa}}, \ and\
  \bibinfo {author} {\bibfnamefont {Y.-i.}\ \bibnamefont {Matsushita}},\ }\href
  {\doibase 10.1063/1.5029535} {\bibfield  {journal} {\bibinfo  {journal} {The
  Journal of Chemical Physics}\ }\textbf {\bibinfo {volume} {148}},\ \bibinfo
  {pages} {224103} (\bibinfo {year} {2018})},\ \Eprint
  {http://arxiv.org/abs/https://doi.org/10.1063/1.5029535}
  {https://doi.org/10.1063/1.5029535} \BibitemShut {NoStop}%
\bibitem [{\citenamefont {{Nishi}}\ \emph {et~al.}(2018)\citenamefont
  {{Nishi}}, \citenamefont {{Kosugi}}, \citenamefont {{Furukawa}},\ and\
  \citenamefont {{Matsushita}}}]{bib:GF_CCSD_and_FCI_for_light_atoms}%
  \BibitemOpen
  \bibfield  {author} {\bibinfo {author} {\bibfnamefont {H.}~\bibnamefont
  {{Nishi}}}, \bibinfo {author} {\bibfnamefont {T.}~\bibnamefont {{Kosugi}}},
  \bibinfo {author} {\bibfnamefont {Y.}~\bibnamefont {{Furukawa}}}, \ and\
  \bibinfo {author} {\bibfnamefont {Y.-i.}\ \bibnamefont {{Matsushita}}},\
  }\href@noop {} {\bibfield  {journal} {\bibinfo  {journal} {ArXiv e-prints}\ }
  (\bibinfo {year} {2018})},\ \Eprint {http://arxiv.org/abs/1803.01512}
  {arXiv:1803.01512 [cond-mat.mtrl-sci]} \BibitemShut {NoStop}%
\bibitem [{\citenamefont {Guan}\ \emph {et~al.}(2009)\citenamefont {Guan},
  \citenamefont {Chen}, \citenamefont {Wang},\ and\ \citenamefont
  {Ma}}]{bib:4493}%
  \BibitemOpen
  \bibfield  {author} {\bibinfo {author} {\bibfnamefont {L.}~\bibnamefont
  {Guan}}, \bibinfo {author} {\bibfnamefont {S.}~\bibnamefont {Chen}}, \bibinfo
  {author} {\bibfnamefont {Y.}~\bibnamefont {Wang}}, \ and\ \bibinfo {author}
  {\bibfnamefont {Z.-Q.}\ \bibnamefont {Ma}},\ }\href {\doibase
  10.1103/PhysRevLett.102.160402} {\bibfield  {journal} {\bibinfo  {journal}
  {Phys. Rev. Lett.}\ }\textbf {\bibinfo {volume} {102}},\ \bibinfo {pages}
  {160402} (\bibinfo {year} {2009})}\BibitemShut {NoStop}%
\bibitem [{\citenamefont {Pecak}\ and\ \citenamefont
  {Sowi\ifmmode~\acute{n}\else \'{n}\fi{}ski}(2016)}]{bib:4494}%
  \BibitemOpen
  \bibfield  {author} {\bibinfo {author} {\bibfnamefont {D.}~\bibnamefont
  {Pecak}}\ and\ \bibinfo {author} {\bibfnamefont {T.}~\bibnamefont
  {Sowi\ifmmode~\acute{n}\else \'{n}\fi{}ski}},\ }\href {\doibase
  10.1103/PhysRevA.94.042118} {\bibfield  {journal} {\bibinfo  {journal} {Phys.
  Rev. A}\ }\textbf {\bibinfo {volume} {94}},\ \bibinfo {pages} {042118}
  (\bibinfo {year} {2016})}\BibitemShut {NoStop}%
\bibitem [{\citenamefont {Nagy}\ and\ \citenamefont
  {Aldazabal}(2011)}]{bib:3964}%
  \BibitemOpen
  \bibfield  {author} {\bibinfo {author} {\bibfnamefont {I.}~\bibnamefont
  {Nagy}}\ and\ \bibinfo {author} {\bibfnamefont {I.}~\bibnamefont
  {Aldazabal}},\ }\href {\doibase 10.1103/PhysRevA.84.032516} {\bibfield
  {journal} {\bibinfo  {journal} {Phys. Rev. A}\ }\textbf {\bibinfo {volume}
  {84}},\ \bibinfo {pages} {032516} (\bibinfo {year} {2011})}\BibitemShut
  {NoStop}%
\bibitem [{\citenamefont {Nagy}\ and\ \citenamefont
  {Aldazabal}(2012)}]{bib:2826}%
  \BibitemOpen
  \bibfield  {author} {\bibinfo {author} {\bibfnamefont {I.}~\bibnamefont
  {Nagy}}\ and\ \bibinfo {author} {\bibfnamefont {I.}~\bibnamefont
  {Aldazabal}},\ }\href {\doibase 10.1103/PhysRevA.85.034501} {\bibfield
  {journal} {\bibinfo  {journal} {Phys. Rev. A}\ }\textbf {\bibinfo {volume}
  {85}},\ \bibinfo {pages} {034501} (\bibinfo {year} {2012})}\BibitemShut
  {NoStop}%
\bibitem [{\citenamefont {Nagy}\ \emph {et~al.}(2012)\citenamefont {Nagy},
  \citenamefont {Aldazabal},\ and\ \citenamefont {Rubio}}]{bib:2825}%
  \BibitemOpen
  \bibfield  {author} {\bibinfo {author} {\bibfnamefont {I.}~\bibnamefont
  {Nagy}}, \bibinfo {author} {\bibfnamefont {I.}~\bibnamefont {Aldazabal}}, \
  and\ \bibinfo {author} {\bibfnamefont {A.}~\bibnamefont {Rubio}},\ }\href
  {\doibase 10.1103/PhysRevA.86.022512} {\bibfield  {journal} {\bibinfo
  {journal} {Phys. Rev. A}\ }\textbf {\bibinfo {volume} {86}},\ \bibinfo
  {pages} {022512} (\bibinfo {year} {2012})}\BibitemShut {NoStop}%
\bibitem [{\citenamefont {Kosugi}\ and\ \citenamefont
  {Matsushita}(2017)}]{bib:4276}%
  \BibitemOpen
  \bibfield  {author} {\bibinfo {author} {\bibfnamefont {T.}~\bibnamefont
  {Kosugi}}\ and\ \bibinfo {author} {\bibfnamefont {Y.-i.}\ \bibnamefont
  {Matsushita}},\ }\href {\doibase 10.1063/1.4994720} {\bibfield  {journal}
  {\bibinfo  {journal} {The Journal of Chemical Physics}\ }\textbf {\bibinfo
  {volume} {147}},\ \bibinfo {pages} {114105} (\bibinfo {year} {2017})},\
  \Eprint {http://arxiv.org/abs/https://doi.org/10.1063/1.4994720}
  {https://doi.org/10.1063/1.4994720} \BibitemShut {NoStop}%
\bibitem [{\citenamefont {Calogero}(1969)}]{bib:3935}%
  \BibitemOpen
  \bibfield  {author} {\bibinfo {author} {\bibfnamefont {F.}~\bibnamefont
  {Calogero}},\ }\href {\doibase 10.1063/1.1664820} {\bibfield  {journal}
  {\bibinfo  {journal} {Journal of Mathematical Physics}\ }\textbf {\bibinfo
  {volume} {10}},\ \bibinfo {pages} {2191} (\bibinfo {year} {1969})},\ \Eprint
  {http://arxiv.org/abs/https://doi.org/10.1063/1.1664820}
  {https://doi.org/10.1063/1.1664820} \BibitemShut {NoStop}%
\bibitem [{\citenamefont {Taut}\ \emph {et~al.}(2003)\citenamefont {Taut},
  \citenamefont {Pernal}, \citenamefont {Cioslowski},\ and\ \citenamefont
  {Staemmler}}]{bib:2028}%
  \BibitemOpen
  \bibfield  {author} {\bibinfo {author} {\bibfnamefont {M.}~\bibnamefont
  {Taut}}, \bibinfo {author} {\bibfnamefont {K.}~\bibnamefont {Pernal}},
  \bibinfo {author} {\bibfnamefont {J.}~\bibnamefont {Cioslowski}}, \ and\
  \bibinfo {author} {\bibfnamefont {V.}~\bibnamefont {Staemmler}},\ }\href
  {\doibase 10.1063/1.1542874} {\bibfield  {journal} {\bibinfo  {journal} {The
  Journal of Chemical Physics}\ }\textbf {\bibinfo {volume} {118}},\ \bibinfo
  {pages} {4861} (\bibinfo {year} {2003})},\ \Eprint
  {http://arxiv.org/abs/https://doi.org/10.1063/1.1542874}
  {https://doi.org/10.1063/1.1542874} \BibitemShut {NoStop}%
\bibitem [{\citenamefont {Laufer}\ and\ \citenamefont
  {Krieger}(1986)}]{bib:2819}%
  \BibitemOpen
  \bibfield  {author} {\bibinfo {author} {\bibfnamefont {P.~M.}\ \bibnamefont
  {Laufer}}\ and\ \bibinfo {author} {\bibfnamefont {J.~B.}\ \bibnamefont
  {Krieger}},\ }\href {\doibase 10.1103/PhysRevA.33.1480} {\bibfield  {journal}
  {\bibinfo  {journal} {Phys. Rev. A}\ }\textbf {\bibinfo {volume} {33}},\
  \bibinfo {pages} {1480} (\bibinfo {year} {1986})}\BibitemShut {NoStop}%
\bibitem [{\citenamefont {Kais}\ \emph {et~al.}(1993)\citenamefont {Kais},
  \citenamefont {Herschbach}, \citenamefont {Handy}, \citenamefont {Murray},\
  and\ \citenamefont {Laming}}]{bib:4476}%
  \BibitemOpen
  \bibfield  {author} {\bibinfo {author} {\bibfnamefont {S.}~\bibnamefont
  {Kais}}, \bibinfo {author} {\bibfnamefont {D.~R.}\ \bibnamefont
  {Herschbach}}, \bibinfo {author} {\bibfnamefont {N.~C.}\ \bibnamefont
  {Handy}}, \bibinfo {author} {\bibfnamefont {C.~W.}\ \bibnamefont {Murray}}, \
  and\ \bibinfo {author} {\bibfnamefont {G.~J.}\ \bibnamefont {Laming}},\
  }\href {\doibase 10.1063/1.465765} {\bibfield  {journal} {\bibinfo  {journal}
  {The Journal of Chemical Physics}\ }\textbf {\bibinfo {volume} {99}},\
  \bibinfo {pages} {417} (\bibinfo {year} {1993})},\ \Eprint
  {http://arxiv.org/abs/https://doi.org/10.1063/1.465765}
  {https://doi.org/10.1063/1.465765} \BibitemShut {NoStop}%
\bibitem [{\citenamefont {Qian}\ and\ \citenamefont {Sahni}(1998)}]{bib:2046}%
  \BibitemOpen
  \bibfield  {author} {\bibinfo {author} {\bibfnamefont {Z.}~\bibnamefont
  {Qian}}\ and\ \bibinfo {author} {\bibfnamefont {V.}~\bibnamefont {Sahni}},\
  }\href {\doibase 10.1103/PhysRevA.57.2527} {\bibfield  {journal} {\bibinfo
  {journal} {Phys. Rev. A}\ }\textbf {\bibinfo {volume} {57}},\ \bibinfo
  {pages} {2527} (\bibinfo {year} {1998})}\BibitemShut {NoStop}%
\bibitem [{\citenamefont {Filippi}\ \emph {et~al.}(1994)\citenamefont
  {Filippi}, \citenamefont {Umrigar},\ and\ \citenamefont {Taut}}]{bib:4477}%
  \BibitemOpen
  \bibfield  {author} {\bibinfo {author} {\bibfnamefont {C.}~\bibnamefont
  {Filippi}}, \bibinfo {author} {\bibfnamefont {C.~J.}\ \bibnamefont
  {Umrigar}}, \ and\ \bibinfo {author} {\bibfnamefont {M.}~\bibnamefont
  {Taut}},\ }\href {\doibase 10.1063/1.466658} {\bibfield  {journal} {\bibinfo
  {journal} {The Journal of Chemical Physics}\ }\textbf {\bibinfo {volume}
  {100}},\ \bibinfo {pages} {1290} (\bibinfo {year} {1994})},\ \Eprint
  {http://arxiv.org/abs/https://doi.org/10.1063/1.466658}
  {https://doi.org/10.1063/1.466658} \BibitemShut {NoStop}%
\bibitem [{\citenamefont {Taut}\ \emph {et~al.}(1998)\citenamefont {Taut},
  \citenamefont {Ernst},\ and\ \citenamefont {Eschrig}}]{bib:903}%
  \BibitemOpen
  \bibfield  {author} {\bibinfo {author} {\bibfnamefont {M.}~\bibnamefont
  {Taut}}, \bibinfo {author} {\bibfnamefont {A.}~\bibnamefont {Ernst}}, \ and\
  \bibinfo {author} {\bibfnamefont {H.}~\bibnamefont {Eschrig}},\ }\href
  {http://stacks.iop.org/0953-4075/31/i=12/a=007} {\bibfield  {journal}
  {\bibinfo  {journal} {Journal of Physics B: Atomic, Molecular and Optical
  Physics}\ }\textbf {\bibinfo {volume} {31}},\ \bibinfo {pages} {2689}
  (\bibinfo {year} {1998})}\BibitemShut {NoStop}%
\bibitem [{\citenamefont {Ivanov}\ \emph {et~al.}(1999)\citenamefont {Ivanov},
  \citenamefont {Burke},\ and\ \citenamefont {Levy}}]{bib:4478}%
  \BibitemOpen
  \bibfield  {author} {\bibinfo {author} {\bibfnamefont {S.}~\bibnamefont
  {Ivanov}}, \bibinfo {author} {\bibfnamefont {K.}~\bibnamefont {Burke}}, \
  and\ \bibinfo {author} {\bibfnamefont {M.}~\bibnamefont {Levy}},\ }\href
  {\doibase 10.1063/1.478959} {\bibfield  {journal} {\bibinfo  {journal} {The
  Journal of Chemical Physics}\ }\textbf {\bibinfo {volume} {110}},\ \bibinfo
  {pages} {10262} (\bibinfo {year} {1999})},\ \Eprint
  {http://arxiv.org/abs/https://doi.org/10.1063/1.478959}
  {https://doi.org/10.1063/1.478959} \BibitemShut {NoStop}%
\bibitem [{\citenamefont {Tempel}\ \emph {et~al.}(2009)\citenamefont {Tempel},
  \citenamefont {Martinez},\ and\ \citenamefont {Maitra}}]{bib:3919}%
  \BibitemOpen
  \bibfield  {author} {\bibinfo {author} {\bibfnamefont {D.~G.}\ \bibnamefont
  {Tempel}}, \bibinfo {author} {\bibfnamefont {T.~J.}\ \bibnamefont
  {Martinez}}, \ and\ \bibinfo {author} {\bibfnamefont {N.~T.}\ \bibnamefont
  {Maitra}},\ }\href {\doibase 10.1021/ct800535c} {\bibfield  {journal}
  {\bibinfo  {journal} {Journal of Chemical Theory and Computation}\ }\textbf
  {\bibinfo {volume} {5}},\ \bibinfo {pages} {770} (\bibinfo {year} {2009})},\
  \bibinfo {note} {pMID: 26609582},\ \Eprint
  {http://arxiv.org/abs/https://doi.org/10.1021/ct800535c}
  {https://doi.org/10.1021/ct800535c} \BibitemShut {NoStop}%
\bibitem [{\citenamefont {Capone}\ \emph {et~al.}(2007)\citenamefont {Capone},
  \citenamefont {de' Medici},\ and\ \citenamefont {Georges}}]{bib:3066}%
  \BibitemOpen
  \bibfield  {author} {\bibinfo {author} {\bibfnamefont {M.}~\bibnamefont
  {Capone}}, \bibinfo {author} {\bibfnamefont {L.}~\bibnamefont {de' Medici}},
  \ and\ \bibinfo {author} {\bibfnamefont {A.}~\bibnamefont {Georges}},\ }\href
  {\doibase 10.1103/PhysRevB.76.245116} {\bibfield  {journal} {\bibinfo
  {journal} {Phys. Rev. B}\ }\textbf {\bibinfo {volume} {76}},\ \bibinfo
  {pages} {245116} (\bibinfo {year} {2007})}\BibitemShut {NoStop}%
\bibitem [{\citenamefont {Glasser}\ and\ \citenamefont
  {Nieto}(2015)}]{bib:3963}%
  \BibitemOpen
  \bibfield  {author} {\bibinfo {author} {\bibfnamefont {M.}~\bibnamefont
  {Glasser}}\ and\ \bibinfo {author} {\bibfnamefont {L.}~\bibnamefont
  {Nieto}},\ }\href {\doibase 10.1139/cjp-2015-0356} {\bibfield  {journal}
  {\bibinfo  {journal} {Canadian Journal of Physics}\ }\textbf {\bibinfo
  {volume} {93}},\ \bibinfo {pages} {1588} (\bibinfo {year} {2015})},\ \Eprint
  {http://arxiv.org/abs/https://doi.org/10.1139/cjp-2015-0356}
  {https://doi.org/10.1139/cjp-2015-0356} \BibitemShut {NoStop}%
\bibitem [{\citenamefont {Temme}(2000)}]{bib:3956}%
  \BibitemOpen
  \bibfield  {author} {\bibinfo {author} {\bibfnamefont {N.~M.}\ \bibnamefont
  {Temme}},\ }\href {\doibase https://doi.org/10.1016/S0377-0427(00)00347-2}
  {\bibfield  {journal} {\bibinfo  {journal} {Journal of Computational and
  Applied Mathematics}\ }\textbf {\bibinfo {volume} {121}},\ \bibinfo {pages}
  {221 } (\bibinfo {year} {2000})}\BibitemShut {NoStop}%
\bibitem [{\citenamefont {Helbig}\ \emph {et~al.}(2009)\citenamefont {Helbig},
  \citenamefont {Tokatly},\ and\ \citenamefont {Rubio}}]{bib:3915}%
  \BibitemOpen
  \bibfield  {author} {\bibinfo {author} {\bibfnamefont {N.}~\bibnamefont
  {Helbig}}, \bibinfo {author} {\bibfnamefont {I.~V.}\ \bibnamefont {Tokatly}},
  \ and\ \bibinfo {author} {\bibfnamefont {A.}~\bibnamefont {Rubio}},\ }\href
  {\doibase 10.1063/1.3271392} {\bibfield  {journal} {\bibinfo  {journal} {The
  Journal of Chemical Physics}\ }\textbf {\bibinfo {volume} {131}},\ \bibinfo
  {pages} {224105} (\bibinfo {year} {2009})},\ \Eprint
  {http://arxiv.org/abs/https://doi.org/10.1063/1.3271392}
  {https://doi.org/10.1063/1.3271392} \BibitemShut {NoStop}%
\bibitem [{\citenamefont {Ben\'{\i}tez}\ and\ \citenamefont
  {Proetto}(2016)}]{bib:3914}%
  \BibitemOpen
  \bibfield  {author} {\bibinfo {author} {\bibfnamefont {A.}~\bibnamefont
  {Ben\'{\i}tez}}\ and\ \bibinfo {author} {\bibfnamefont {C.~R.}\ \bibnamefont
  {Proetto}},\ }\href {\doibase 10.1103/PhysRevA.94.052506} {\bibfield
  {journal} {\bibinfo  {journal} {Phys. Rev. A}\ }\textbf {\bibinfo {volume}
  {94}},\ \bibinfo {pages} {052506} (\bibinfo {year} {2016})}\BibitemShut
  {NoStop}%
\bibitem [{\citenamefont {Nishida}(2018)}]{bib:bosons1}%
  \BibitemOpen
  \bibfield  {author} {\bibinfo {author} {\bibfnamefont {Y.}~\bibnamefont
  {Nishida}},\ }\href {\doibase 10.1103/PhysRevA.97.061603} {\bibfield
  {journal} {\bibinfo  {journal} {Phys. Rev. A}\ }\textbf {\bibinfo {volume}
  {97}},\ \bibinfo {pages} {061603} (\bibinfo {year} {2018})}\BibitemShut
  {NoStop}%
\bibitem [{\citenamefont {Pricoupenko}(2018)}]{bib:bosons2}%
  \BibitemOpen
  \bibfield  {author} {\bibinfo {author} {\bibfnamefont {L.}~\bibnamefont
  {Pricoupenko}},\ }\href {\doibase 10.1103/PhysRevA.97.061604} {\bibfield
  {journal} {\bibinfo  {journal} {Phys. Rev. A}\ }\textbf {\bibinfo {volume}
  {97}},\ \bibinfo {pages} {061604} (\bibinfo {year} {2018})}\BibitemShut
  {NoStop}%
\bibitem [{\citenamefont {Guijarro}\ \emph {et~al.}(2018)\citenamefont
  {Guijarro}, \citenamefont {Pricoupenko}, \citenamefont {Astrakharchik},
  \citenamefont {Boronat},\ and\ \citenamefont {Petrov}}]{bib:bosons3}%
  \BibitemOpen
  \bibfield  {author} {\bibinfo {author} {\bibfnamefont {G.}~\bibnamefont
  {Guijarro}}, \bibinfo {author} {\bibfnamefont {A.}~\bibnamefont
  {Pricoupenko}}, \bibinfo {author} {\bibfnamefont {G.~E.}\ \bibnamefont
  {Astrakharchik}}, \bibinfo {author} {\bibfnamefont {J.}~\bibnamefont
  {Boronat}}, \ and\ \bibinfo {author} {\bibfnamefont {D.~S.}\ \bibnamefont
  {Petrov}},\ }\href {\doibase 10.1103/PhysRevA.97.061605} {\bibfield
  {journal} {\bibinfo  {journal} {Phys. Rev. A}\ }\textbf {\bibinfo {volume}
  {97}},\ \bibinfo {pages} {061605} (\bibinfo {year} {2018})}\BibitemShut
  {NoStop}%
\bibitem [{\citenamefont {Stefanucci}\ and\ \citenamefont {van
  Leeuwen}(2013)}]{stefanucci2013nonequilibrium}%
  \BibitemOpen
  \bibfield  {author} {\bibinfo {author} {\bibfnamefont {G.}~\bibnamefont
  {Stefanucci}}\ and\ \bibinfo {author} {\bibfnamefont {R.}~\bibnamefont {van
  Leeuwen}},\ }\href {https://books.google.co.jp/books?id=6GsrjPFXLDYC} {\emph
  {\bibinfo {title} {Nonequilibrium Many-Body Theory of Quantum Systems}}}\
  (\bibinfo  {publisher} {Cambridge University Press},\ \bibinfo {year}
  {2013})\BibitemShut {NoStop}%
\end{thebibliography}%

\end{document}